\documentstyle[eqsecnum,prb,aps]{revtex}
\input epsf
\epsfverbosetrue

\newcommand{\resetcounters}{
        \setcounter{equation}{0}
        \setcounter{figure}{0}
        \setcounter{table}{0} }

\begin{document}

\draft

\title{ Effects of Phonons and Nuclear Spins on the Tunneling of a Domain Wall}

\author{$\mbox{M. Dub\'e}^{1,3,4} \, \mbox{and} \,
\mbox{P. C. E. Stamp}^{1,2}$}

\address{ $^{1}$Physics and Astronomy Department and
 $^{2}$Canadian Institute for Advanced Research, \\
University of British Columbia, 6224 Agricultural
rd., Vancouver, BC, Canada V6T 1Z1 \\
$^{3}$Helsinki Institute of Physics, P.O. Box 9 (Siltavuorenpenger 20 C),
FIN-00014, \\ University of Helsinki, Finland \\
$^{4}$Laboratory of Physics,  P.O. Box 1000, FIN-02150 HUT,  \\
Helsinki University of Technology, Espoo, Finland}
\maketitle

\begin{abstract}

We consider the quantum dynamics of a magnetic domain wall at low
temperatures, where dissipative
couplings to magnons and electrons are very small. The
wall motion is then determined by its coupling to phonons and nuclear
spins, and any pinning potentials. 

In the absence of nuclear spins
there is a dominant superOhmic  
1-phonon coupling to the wall {\it velocity}, 
plus a strongly $T$-dependent Ohmic coupling to pairs of phonons. 
There is also a 
$T$-independent Ohmic coupling between single phonons and the wall
chirality, which suppresses ``chirality tunneling´´. We calculate the
effect of these couplings on the $T$-dependent tunneling rate of a wall
out of a pinning potential.

Nuclear spins have a very strong and  
hitherto unsuspected influence on domain wall
dynamics, coming from a hyperfine-mediated
coupling to the domain wall position. 
For $k_B T \gg \omega_{0}$ this coupling yields a spatially random
potential, 
fluctuating at a rate governed by the nuclear $T_2$.
When $k_B T \ll \omega_{0}$, the hyperfine potential fluctuates around a
{\it linear} binding potential. 

The wall dynamics is influenced by the
fluctuations of this potential, ie., by the
nuclear spin dynamics. Wall
tunneling can occur when
fluctuations open
an occasional ``tunneling window''.
This changes the crossover to tunneling
and also causes a slow "wandering", in time, of the energy levels
associated with domain wall motion inside the pinning potential. 
This effect is fairly weak in $Ni$- and $Fe$- based magnets, and 
we give an approximate treatment of its effect on the tunneling dynamics,
as well as a discussion of the relationship to recent domain wall
tunneling experiments.

\end{abstract}

\pacs{PACS numbers:     }

\section{Introduction}
\resetcounters

The quantum dynamics of mesoscopic or even macroscopic magnetic solitons
has been the subject of considerable interest in the last few years.
There are several reasons for this. One is the rapid development of the
new field of nanomagnetism \cite{PhysToday}, 
in which very small magnetic wires, films,
and magnetic macromolecules \cite{Gatteschi} 
(as well as molecular chains) can be prepared. 
In many cases one expects that solitons such as domain walls, or perhaps
lower-dimensional vortices and skyrmions, will play an important role in
the quantum dynamics of nanomagnets at low temperatures in systems as
diverse as magnetic perovskites, the cuprate superconductors,
or nanomagnetic wires, films, and junctions. The magnetic dynamics
also strongly influence the charge transport in some systems.

A second reason for this interest stems from the continued fascination
with macroscopic quantum phenomena. Previously associated almost
exclusively with superfluids and superconductors, it has become clear in
recent years that magnetic systems may also exhibit macroscopic
quantum phenomena \cite{stamprev,QTM94},
perhaps the most spectacular being
the possible quantum tunneling of macroscopic domain walls in
ferromagnets \cite{stamprev,stampprl}. 
There now exists some experimental support for
tunneling of {\it single} domain walls in Ni wires 
\cite{giordtun,giordesc,giordres,KHong},
as well as many earlier experiments on multi-wall systems 
\cite{Uehara87,Paulsen91,Zhang92}. 
The tunneling dynamics of large flexible domain
walls poses many interesting theoretical problems, particularly
concerning the role played by the various environmental degrees of
freedom, which couple to the wall in various ways.

This is the main subject of this paper, which deals with the hitherto
neglected effects of phonons and nuclear spins on magnetic domain wall
tunneling (a limited discussion of phonon effects was given previously 
\cite{stamprev,stampprl}, but was very 
incomplete). We find that the effect of the nuclear spins in particular  
can be quite spectacular, and can completely contradict the 
picture of quantum domain
wall dynamics that has evolved over the last few years. We
derive a new scenario, in which 
magnetic domain wall dynamics 
is influenced at low temperatures by the effective
coupling of the wall to a random (in space) ``hyperfine field'', which
also fluctuates in time. Even in Ni wires (where these effects
are far weaker than in other magnets), this fluctuating
hyperfine field will be shown to have some influence on wall
tunneling. The phonons are found to have a much weaker effect, 
but one nevertheless important to evaluate when one considers the
finite-T tunneling rate 
(particularly
in isotopically purified systems, where phonons entirely determine the
low-$T$ dissipative dynamics, if the system is pure and insulating).
We shall also see that phonons couple very strongly to the domain wall
chirality, and thereby suppress chirality tunneling. 

In the remainder of this section we give a brief explanation of previous
work in this field, and note how the present paper fits into this.

\subsection{Macroscopic Quantum Phenomena in Magnets}

The history of macroscopic quantum phenomena goes back to the 1930's 
(Meissner effect, superfluid fountain effect, capillary superflow, etc),
and has until recently been associated almost exclusively with
superconductors and superfluids. In recent years considerable attention
has focused on Caldeira and Leggett's proposal that SQUIDs and related
systems might show macroscopic flux tunneling \cite{leggettA}. Subsequent
experiments \cite{voss,washburn,clarke1,lukens1,clarke2} confirmed their
work quantitatively, and this is still an active field \cite{lukens2}.

It is obviously important to find other systems
in which quantum macroscopic phenomena occur. It should be
stressed that this is not just a question of energetics but
also involves the ``environmental decoherence'' which almost always
destroys quantum phase correlations at all but the microscopic level. 
Consider, eg.,  the possibility of ``macroscopic
quantum tunneling'' (MQT) just mentioned. Many authors have suggested
MQT after having noted the existence of some small tunneling barrier
acting on some large collective mode - there is a long list of
candidates for such MQT. The list includes $^{4}\mbox{He}$ superfluid
\cite{Vinen,Maris}and $^{3}\mbox{He}$ superfluid \cite{Bailin}, 
superconductors \cite{leggettA,Duan}
ferroelectrics \cite{Melo}, 
charge density waves (CDW) tunneling in
1-dimensional systems \cite{Bardeen}, MQT and tunneling nucleation of
dislocations, vacancies and impurities in quantum solids
\cite{Andreev}, quantum diffusion of such objects in quantum
solids \cite{KagProk}, as well as in normal \cite{ProkFLiq} and 
superfluid \cite{ProkHeB} $^{3} \mbox{He}$. 
To this list we may also
add more general studies of quantum nucleation 
\cite{Lifshitz} as well as 
of large scale tunneling in cosmology \cite{Coleman}.
This list in certainly not complete. Finally there
are the various kinds of macroscopic quantum phenomena, including MQT,
which have been suggested for magnetic systems, to which we come below.

Despite the length of this list, in
few cases have these suggestions been confirmed by experiments. 
There are two main reasons for this. First, the 
the surrounding ``quantum environment''  
usually strongly suppresses MQT (and completely suppresses ``macroscopic
quantum coherence'' (MQT)). 
Consequently, as
emphasised by Leggett et. al
\cite{leggettA,leggettprb,leggettRMP,leggettVar}
any discussion of MQT which
ignores the environmental effects is usually of of academic interest only.

A second reason is simply that most 
physical materials contain defects, impurities and other imperfections.
The coupling of these to the collective coordinates involved in MQT
is often much larger than the very small tunneling barriers.
Since such external effects are usually unknown, this makes 
experimental tests very difficult.

So far, the only case where really indisputable evidence can be given  
for the observation of MQT is in superconductors (in SQUIDs and other
Josephson devices \cite{clarke2}).
This work 
has been widely reviewed \cite{clarke2,leggettVar}. In
magnetic systems, suggestions that tunneling phenomena might exist go
back over 40 years \cite{Weil,bean}. 
These suggestions typically
involved some 1-dimensional phenomenological potential (a ``pinning 
potential'' for domain walls, an anisotropy potential for monodomain
grains) to which a naive WKB analysis 
was applied. No microscopic theory was attempted, and the experimental
evidence lacked any clear theoretical basis.

However, the picture that has emerged in recent years is very different from
this, for 2 main reasons. The first is 
the realisation that a standard WKB analysis is
inappropriate to spin systems, in which a constraint typically exists -
in small particles or macromolecules at low T 
the magnetisation density ${\bf |M|}$ is roughly constant (although not
necessarily in domain walls, if  
multi-magnon excitations are important \cite{stampprl}). 
This constraint leads to a
kinetic energy which is linear in time derivatives (not quadratic). Thus
a correct solution for the problem of tunneling of a spin was not given
until 1986, by van Hemmen and Suto
\cite{vanhemmen1,vanhemmen2,scharf} (and independently for a
particular case by Enz and Schilling \cite{enz1}). Both of these
analyses were semi-classical. A key role in nanomagnetic dynamics is also
played by the
``Kramers/Haldane phase'', which differs for integer and half-integer
spins; this has no analogue in superconductors 
\cite{vanhemmen2,LossPRL,Henley}.

The second reason for the revised picture is that once
microscopic theories were attempted of 
domain wall tunneling, or of tunneling in  
nanomagnets, it became obvious that yet again, much of the physics was
in the coupling to the background quantum environment. Some of this
physics is very different from that in superconductors. In particular,
the coupling 
to nuclear spins is very strong, can play a major role in
controlling the quantum dynamics of mesoscopic or nanoscopic magnets,
and cannot obviously be understood in terms of an ``oscillator bath'' model
of the environment
\cite{prostamp1,stamp1,prostamp2,prostamp3,prostamp4,dauriac}. 
Extensive theory
has been done in an effort to understand the way in which the environment
determines nanomagnetic dynamics, both in particles and molecules
\cite{stamprev,prostamp1,prostamp4,GargKim,Politi1}
and in domain wall dynamics (see below).

Since this theoretical work was done, strong experimental evidence has
emerged for tunneling phenomena in mesoscopic magnetic systems at low
temperature. This includes many experiments on multi-grain systems 
\cite{multigrain} and
on systems with multiple domains \cite{multiwall}; 
evidence has been claimed both for
magnetisation tunneling in monodomain systems and for domain wall
tunneling in multi-domain systems. There has also been a claim for the
observation of coherent tunneling in large ferritin molecules
\cite{awshalom}, which has
been disputed by various authors \cite{rebuke}. A very interesting 
set of experiments by Giordano et. al. 
\cite{giordtun,giordesc,giordres,KHong}
reports strong evidence for tunneling of single domain walls in Ni
wires, to which we will return below. Finally, a whole series of
experiments on $\mbox{Mn}_{12}$-acetate molecular crystals gives strong
evidence for thermally activated resonant tunneling in this system 
\cite{Mn12}. 

\subsection{Domain Wall Tunneling}

The principal results to emerge from the first microscopic analyses of
domain wall tunneling have been: 

(a) The effective pinning potential for a wall is almost independent of
the nature of the pinning centre, provided the spatial extent of the
pinning centre (usually some kind of defect) is much smaller than the
wall width $\lambda$; the shape of the defect is also 
not important in this
case \cite{stamprev}. The pinning potential is of the form 
$V(Q) = -V_{0} \mbox{sech}^{2}(Q/\lambda)$, where $V_{0}$ is positive.
Subsequent analyses of the so-called ``domain wall junction'' rely on
this result, which is important for any comparison with experiment,
simply because the exact nature of the pinning centre is not known. The
constant $V_{0}$ can be determined experimentally in various ways (eg.,
from low frequency ``wall rocking'' absorption measurements, or, less
reliably, form a determination of the wall escape characteristics at
higher T).

(b) For an insulating magnet, the principal contribution to wall
dissipation at temperatures above the magnon gap $\Delta$ will come from
multi-magnon emission by the wall \cite{stampprl}. 
However if $kT \ll \Delta$, the main
magnon contribution to the dissipation will come from processes
involving 2 bulk magnons and one wall, or ``Winter'', magnon; this
3-magnon process gives an Ohmic contribution to the wall dissipation 
coefficient $\eta$, going like 
$\eta_{3} \sim (kT/\Delta) \exp (-\Delta/kT)$. Higher magnon processes
give contributions 
$\eta_{m} \sim (kT/\Delta)^{m-2} \exp (-\Delta/kT)$, and are thus
subdominant. 2-magnon processes give a superOhmic contribution
to the wall dissipation which is negligible compared to the Ohmic
contribution at low energy or low wall velocity. 

(c) When $kT \ll \Delta$, so that the magnons have negligible effects on
the wall dynamics, an analysis has been given of  
2 other small contributions to dissipation \cite{stamprev}:
a coupling between the wall and Winter magnons, mediated by defects;
and also a quite negligible  
coupling to photons. 
The influence of the defect-mediated coupling has
recently been re-examined by Leduc and Stamp \cite{leduc}, 
with the conclusion that
it can usually be neglected in practically relevant calculations. 

(d) For a conducting magnetic system, a  
thorough analysis
by Tatara and Fukuyama \cite{tatara2} showed that both the charge 
fluctuation (eddy
current) and spin fluctuation contributions to the wall dissipation will
be negligible unless the wall thickness is not much larger than
the lattice separation between magnetic ions (even for
$\mbox{SmCo}_{5}$, where $\lambda = 12 \AA$, the electronic contribution
gave a correction $< 1 \%$ to the tunneling exponent).

(e) The phonon contribution to the dissipation of wall motion was
analysed \cite{stamprev,stampprl} using the method of 
Wada and Schrieffer \cite{wada} to give a small $T^{3}$
contribution to the wall diffusion constant. This analysis was 
incomplete, since it did not include all one-phonon
contributions to the dissipation; moreover, it did not include higher
non-linear couplings, such as 2-phonon couplings. Since the 1-phonon
coupling is superOhmic, experience with magnon couplings suggests one
should go to higher order to look for Ohmic contributions to the wall
dissipation. This is one of the main tasks of the present paper. 

(f) One can also consider the possibility of ``chirality tunneling'' of
the domain wall chirality \cite{braun2,tatara1}, or even the coherent
Bloch dynamics of walls in a periodic potential \cite{braun1}. 
Until now no attempt has been made to see whether these  
possibilities would survive the coupling to the environment. This issue
is also addressed herein; we find that they do not 
survive.

Amongst the various experimental searches for macroscopic domain wall 
tunneling, perhaps the most 
dramatic results are those of Giordano et. al.
\cite{giordtun,giordesc,giordres,KHong}.
These results are in fairly good agreement  
with the theory, except that the experimental result for the
quantum/classical crossover temperature (between tunneling and activated
escape) is over an order of magnitude higher than the theoretical
prediction \cite{stampprl}.

In this paper we will show that yet another 
modification of our understanding of the low-$T$ dynamics of domain
walls (and other magnetic solitons) is necessary. This is because
of the inevitable coupling of the wall to the background nuclear spins,
via hyperfine interactions. We shall see that the  
longitudinal hyperfine coupling energy
$ \sum_{k} \omega_{k} s_{k}^{z} I_{k}^{z}$, summed over all
nuclear spins ${\bf I}_{k}$ 
and electronic spins ${\bf s}_{k}$) in the system, can be very large. 
The
transverse hyperfine coupling 
$ \sum_{k} \omega_{k} ( s_{k}^{+} I_{k}^{-} + 
 s_{k}^{-} I_{k}^{+})$ causes irreversible transitions in the
nuclear spin bath every time the nanomagnetic system changes its state.
This changes the energetics of the tunneling, and also causes
``topological decoherence'' and dissipation. These names are imported from
the study of environmental ``spin bath'' effects on nanomagnetic
tunneling
\cite{prostamp1,prostamp3}, where similar physics applies. 

We will not try here to give a complete analysis of
nuclear spins effects on domain wall dynamics - this would be a very
lengthy task. We concentrate instead on the 
static longitudinal hyperfine coupling. 
The main conclusion, developed in
detail below, is that domain wall tunneling is rendered much more
complex by the coupling to nuclear spins and to phonons. The
longitudinal hyperfine field is random in space, and fluctuates in time
at a rate governed principally by $T_2$ relaxation. It can be very strong,
and typically pins the domain wall. Tunneling can occur when temporal
fluctuations in this potential open up ``energy windows'' for brief
periods of time. The phonons also allow
inelastic tunneling processes. Analysis of the 
various couplings to
phonons shows that there are 3 important ones: a linear coupling to the
domain wall velocity, a bilinear Ohmic coupling to the
wall position, and an important linear Ohmic coupling to the wall
chirality, which 
strongly suppresses any chirality tunneling,
even in the absence of nuclear spins. As might have been expected, Bloch
coherence is so fragile that it is destroyed by almost anything, and we
believe it to be practically unobservable.

The plan of this paper is as follows. In the next section we discuss the
various domain walls we will be dealing with in various geometries,
including magnetic wires, magnetic films, and bulk 3-dimensional magnets.
Essential results for the wall dynamics and tunneling in the absence of
environmental effects are given, as well as relevant experimental
numbers.

In section III we derive the effective low-energy interactions which
couple a domain wall to both phonons and nuclear spins. The wall
is coupled simultaneously to an ``oscillator bath'' of phonons,
and a ``spin bath'' of nuclear spins, and a number of general results for
such baths are recalled. 

In section IV an extensive analysis is given of phonon effects on 
domain wall dynamics, for walls in 1, 2, and 3 dimensions, in isotopically
purified magnets (no nuclear spins).
The effective action for the wall is derived, with the phonons integrated
out. These calculations are interesting theoretically, because of the
novelty of some of the couplings.
The dominant couplings to the wall
velocity, the wall position, and the wall chirality are treated
separately.

In section V the problem of wall tunneling is analysed. We first consider
the ``displacement tunneling'' of walls in isotopically purified magnets,
where it is controlled by phonons. We then include the 
nuclear spins, and show that tunneling is then strongly
influenced 
by the dynamic hyperfine potential acting on the wall. For displacement
tunneling 
this changes
the theoretical predictions for the tunneling (in particular
the crossover temperature and the tunneling rate). 
As a corollary to these calculations we
analyse the suppression of both chirality tunneling and of Bloch
coherence.

In the final section we make a few remarks on the relation to experiments
on wall tunneling (particularly those of Giordano et al.).

\section{Domain Walls and Their Energetics}

In this section, we introduce the different kinds of domain wall we wish
to study. Discussion of the
couplings to the environment is reserved
to later sections. We work exclusively with
ferromagnetic systems, at low $T$, and we assume that the domain
wall thickness 
$\lambda$ is sufficiently greater than the lattice separation
$a_{0}$ between spins so that a continuum approximation for the
magnetisation is valid. 
We will be discussing domain walls in non-trivial
geometries, including films and wires, as
well as 3-dimensional systems. 

In all cases, we assume that the magnetisation at position 
${\bf r}$ in the crystal 
is defined as ${\bf M}({\bf r}) = g \mu_{B} \sum_{j}
{\bf s}_{j} \, \delta ({\bf r}-{\bf R}_{j})$, where $\mu_{B}$ is the Bohr
magneton $\mu_{B} = e\hbar /2 m_{e} c$,  $g$ is the Land\'e factor,
and ${\bf R}_{j}$ is the position of the $j^{th}$ lattice site, with  
associated electronic spin ${\bf s}_{j}$ .

There is no easy way to deal with the magnetic structure in bulk
materials and in thin films and wires 
in a unified manner, as the direction of the
magnetisation within the walls is quite different in these two cases.
Instead of considering different values for the magnetisation in the
$(x,y,z)$ directions, we define in each case (bulk and thin films) a new
frame of reference $(x_{1},x_{2},x_{3})$ such that the easy axis is along the
$x_{1}$ direction, with the wall characterised by its position along the
$x_{3}$ axis. 
The magnetisation is written as ${\bf M}=M_{0} \hat{{\bf m}}$ with the unit
vector $\hat{{\bf m}}= (m_{x},m_{y},m_{z}) =
(\cos \phi \sin \theta, \sin \phi \sin \theta, \cos \theta)$
so that $({\bf \nabla m})^{2} = ({\bf \nabla}\theta)^{2} + \sin^{2} \theta
 ({\bf \nabla}\phi)^{2}$.
We will assume that the magnitude $M_{0}$ is constant, 
since all phenomena considered occur at temperatures 
far below the Curie temperature.
We thus use $M_{0} = s \gamma_{g} \hbar/a_{0}^{3}$ where
$\gamma_{g} = g \mu_{B}/\hbar$ is the gyromagnetic factor, $a_{0}$ is the
lattice spacing of the crystal and $s$ is the value of the electronic spin.
We will assume a standard biaxial (easy axis/easy plane)
Hamiltonian, given in continuum approximation by 
\cite{malo}
\begin{equation}
{\cal H} = \frac{1}{2} \int d {\bf r} [ J ({\bf  \nabla M} )^{2}
- K_{\|} (M_{1})^{2} + K_{\bot} (M_{3})^{2} - \frac{\mu_{0}}{2}
( {\bf H}_{dm}+{\bf H}_{e} ) \cdot {\bf M}]
\label{hamiltonian}
\end{equation}
where $({\bf \nabla M})^{2}\equiv (\nabla_{i} M_{j})^{2}$. The easy axis
is represented by the component $M_{1}$ of the magnetisation, and the
easy plane is perpendicular to $M_{3}$. $K_{\|}$ and $K_{\bot}$ are the
anisotropy constants, 
${\bf H}_{e}$ is an external magnetic
field and ${\bf H}_{dm}$ is the demagnetising
field \cite{landau},
the internal field coming from all the magnetic moments.

The action of a the magnetisation also contains a term with no classical
analogue,  the Berry phase factor \cite{berry}, or Wess-Zumino 
term \cite{shapere}, given by
\begin{equation}
S_{WZ} = i \int_{0}^{1/T} \int d {\bf r} \, \dot{\phi}({\bf r},t)
 (1-\cos \theta({\bf r},t))
\end{equation}
so that the total action is $S=S_{WZ} + \int d\tau {\cal H}$. The Wess-Zumino
term is purely imaginary. It essentially corresponds to the total solid angle
traced by the spin configuration for a given trajectory; its
derivation using path-integral or coherent spin state descriptions of
the system is now very well known, following the 1983 work of Haldane 
\cite{auerbach}.

\subsubsection{Bloch Wall in Bulk Materials}

For our study of the tunneling of a Bloch wall, we take the Hamiltonian
\begin{eqnarray}
{\cal H} &=& \int d {\bf r}  [J  ({\bf \nabla m})^{2}
- K_{\|} m_{z}^{2} + K_{\bot} m_{x}^{2}] \nonumber \\
&=& \int d {\bf r} [ J  (({\bf \nabla}\theta)^{2} + \sin^{2} \theta
 ({\bf \nabla}\phi)^{2})
- K_{\|} \cos^{2} \theta + K_{\bot}
\cos^{2} \phi \sin^{2} \theta ]
\end{eqnarray}
representing a ferromagnet with easy $z$ axis
and easy $z-y$ plane. 
In 3-dimensions, $J$ is measured in $\mbox{J/m}$, and the 
anisotropy constants in $\mbox{J/m}^{3}$.

\begin{figure}
\epsfysize=3.5in
\epsfbox[15 450 593 753]{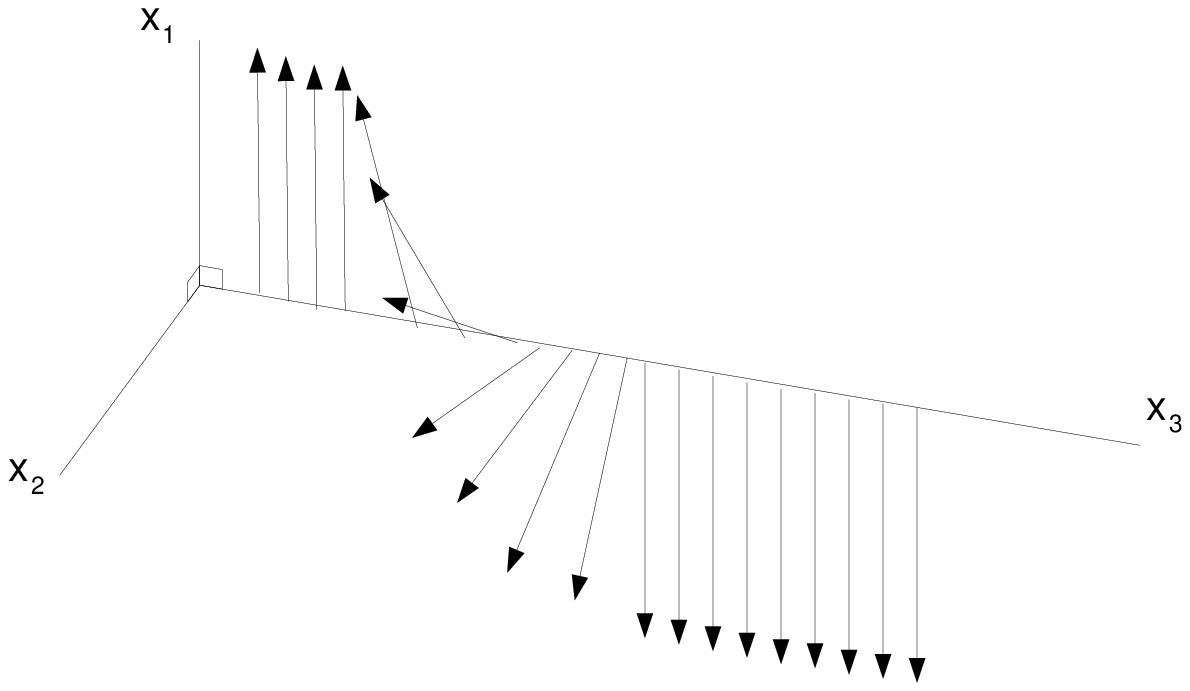}
\caption{The standard model of a Bloch Wall, with axes labelled as in the
text.}
\label{bloch}
\end{figure}

The domain wall corresponding to this Hamiltonian is perpendicular to the
$x$ axis, with the magnetisation rotating in the $z-y$ plane. The wall 
centre is located at a position $Q$ along the $x$ axis.
The new frame of reference is thus $(x_{1},x_{2},x_{3}) =(z,y,x)$.
This is represented in Fig.(\ref{bloch}).

%\begin{figure}[t]
%\epsfysize=3.5in
%\epsfbox[15 450 593 753]{bloch.ps}
%\caption{Model of a Bloch Wall}
%\label{bloch}
%\end{figure}

The components of the magnetisation are given by \cite{malo}:
\begin{eqnarray}
\hat{m}^{B}_{1} &=& C \tanh \left( \frac{x_{3}-Q(t)}{\lambda_{B}}
                    \right) \nonumber \\
\hat{m}^{B}_{2} &=& \chi \left( 1 - \frac{\dot{Q}^{2}(t)}{8 c_{0}^{2}}
                    \right)
\mbox{sech}  \left( \frac{x_{3} -Q(t)}{\lambda_{B}} \right)
\label{magcomp} \\
\hat{m}^{B}_{3} &=& C \frac{\dot{Q}(t)}{2 c_{0}} \mbox{sech} \left(
\frac{x_{3} -Q(t)}{\lambda_{B}} \right) \nonumber
\end{eqnarray}
where the thickness of the wall $\lambda_{B} = (J/K_{\|})^{1/2}$ 
represents the
usual compromise between exchange and anisotropy. The surface energy of this
type of wall is $\sigma_{0}=4(JK_{\|})^{1/2}$. $C = \pm 1$ is the 
``topological charge''
of the wall and $\chi = \pm 1$ is the ``chirality''. The 
topological charge corresponds to the
direction along which the wall moves under the application of an external 
magnetic field in direction parallel to the easy axis, while the 
chirality refers to the sense of the rotation of the magnetisation
inside
the wall.
The internal configuration of the wall is such as to 
simultaneously minimize the
anisotropy energy (coming from both $K_{\|}$ and $K_{\bot}$), and the 
exchange energy.
The anisotropy $K_{\bot}$ can originate from the material
itself but also from the configuration of the magnetisation, through the
demagnetisation energy \cite{malo}, ie., 
$K_{\bot} = K_{\bot,i} + \mu_{0} M_{0}^{2}/2$ where $K_{\bot,i}$ is the
anisotropy intrinsic to the material.
A static Bloch wall only rotates in the easy plane. However, as soon as it
moves it creates a demagnetising field which causes the spins to precess and
the appearance of a component of the magnetisation out of the plane, directly
proportional to the wall velocity \cite{malo,odell}.
This picture is valid provided that $\dot{Q}(\tau) << c_{0}$, the Walker
critical velocity,
where \cite{malo}
\begin{equation}
c_{0} = \frac{2 \gamma_{g}}{M_{0}} (J K_{x})^{1/2} \left[ \left[1+
\frac{K_{\bot}}{K_{x}} \right]^{1/2} -1 \right]
\label{vwalker}
\end{equation}
where typically $c_{0} \sim 10^{2} \, \mbox{m/s}$.
The precession of the spins also
causes the appearance of an inertial term, the D\"oring mass, given by
the
ratio of the wall energy with the limiting velocity \cite{malo}:
\begin{equation}
M_{w} = \frac{S_{w} M_{0}^{2}}{\gamma_{g}^{2} (J K_{\|})^{1/2} }
\left[ \frac{1}{( 1+ K_{\bot}/K_{\|})^{1/2}-1} \right]^{2}
\label{mdoring}
\end{equation}
where $S_{w}$ is the surface of the wall.
We assume $K_{\bot} \sim \mu_{0}M^{2}_{0}/2 \gg K_{\|}$,
and in this limit,
$c_{0} = \mu_{0} \gamma_{g} \lambda M_{0}/2$ and the
D\"oring mass reduces to $M_{w} = 2S_{w}/ \mu_{0} \gamma_{g}^{2}
\lambda$.

All of the above formulae for the Bloch wall are well known, and should
be treated as phenomenological; they are given in terms of microscopic
parameters, but should not be treated as the result of a microscopic
derivation. In fact such a derivation would certainly give somewhat
different results in the long-wavelength limit appropriate to the
macroscopic dynamics of the wall, since integrating out the short
wavelength fluctuations in ${\bf M}({\bf r},t)$ will renormalise all
parameters. We do not bother to do this for 2 reasons, viz.,

(a) It is obvious that the renormalising influence of the
high-frequency fluctuations will be strongly suppressed by the
long-range demagnetisation field 

(b) The only important parameter in any experimental tests of the
theory will be the Doring mass, and this will have to measured anyway
(by, eg., looking at the fluctuation frequency of the wall in a pinning
potential); the measured mass will of course then already incorporate
the renormalisations.
 
Because of (a) we expect that the expressions in (\ref{vwalker}) and 
(\ref{mdoring}) 
to be fairly accurate anyway, and we will use them when necessary.

If defects are present in the sample, they can pin the wall. We 
will assume that the radius $R_{d}$ corresponding to the defect volume 
is much smaller than $\lambda_{B}$, the domain wall width.
The wall is then pinned by a
potential of form \cite{stampprl}
\begin{equation}
V(Q) = - V_{0} \mbox{sech}^{2} (Q/\lambda_{B})
\label{pinpot}
\end{equation}
with $V_{0}$ proportional to the volume of the defect
\cite{stampprl}.
We further assume that there
is a very small concentration of defects, so there is only 1
important pinning centre for the wall. This would correspond to an ideal
experimental situation. We also assume that the wall is flat and
that it remains flat during the tunneling process. This approximation is
justified by the energy associated with the curvature of the wall, to
which there are two contributions. The first is the surface energy
$\sigma_{0}$ of the wall. A curved wall has a larger surface, and thus a
larger energy. Second, 
and much more important, a curvature in the $x_{1}-x_{3}$ plane 
creates strong long-range demagnetisation fields, 
which rapidly increase
the wall energy. A complete treatment of these effects is quite
complicated \cite{malo}, but for small
pinning energies, the radius of curvature of
the wall is much larger than $\lambda$, and it is easily shown that weak
curvature has very little effect on wall tunneling
\cite{stamprev,stampprb}. To consider the effect of dissipation, we can thus 
use a flat wall.

The application of an external  magnetic field ${\bf H}_{e}$ in the
direction
of the easy axis couples to the magnetisation to give a potential term
linear in $Q$. Including the inertial mass, we then write a ``bare''
Hamiltonian (ie., neglecting the environment) for the wall \cite{stampprl}
\begin{equation}
H_{w} = \frac{1}{2} M_{w} \dot{Q}^{2} - V(Q) - 2 S_{w} \mu_{0} M_{0}
H_{e} Q
\end{equation}
where we put the chirality $C = 1$ for brevity.

The tunneling rate of the wall can now be evaluated by
standard instanton techniques. For a field sufficiently close to the
coercive field, the
``pinning potential plus field potential'' reduces to
a quadratic plus cubic potential, whose barrier
height is controlled by $\epsilon=1-H_{e}/H_{c}$, $H_{c}$ being the
coercive field, given as
\begin{equation}
\mu_{0} H_{c} = \frac{2}{3\sqrt{3}} \frac{V_{0}}{\lambda S_{w} M_{0}}
\label{coer}
\end{equation}
while the energy barrier is
\begin{equation}
\tilde{V} (\epsilon) \sim  (\hbar \gamma_{g} \mu_{0} H_{c}) N_{0} \epsilon^{3/2}
\label{barrier}
\end{equation}
where $N_{0}$ is the number of spins in the wall.
The exact value of the coercive field is of course very difficult to obtain
theoretically and should ideally be obtained by a characterisation of
the system in the thermal phase.

The tunneling rate $\Gamma_0$ 
in a cubic plus quadratic potential is well known
\cite{stamprev,stampprl,leggettA}; we simply recall here the results
for the tunneling of the wall in the absence of dissipation:
\begin{equation}
\Gamma_0 = \left[ \frac{30}{\pi} \frac{B(\epsilon)}{\hbar} \right]^{1/2}
\Omega_{0} \; e^{-B_0 (\epsilon)/\hbar}
\end{equation}
with $B_0 (\epsilon)$ the usual WKB tunneling exponent;
\begin{equation}
\frac{1}{\hbar} B_0 (\epsilon)= 
\frac{8}{15} M_{w} \Omega_{0} Q_{0}^{2}
\label{tunnelexp}
\end{equation}
$\Omega_{0}$ the
oscillation frequency of the wall in the potential;
\begin{equation}
\Omega_{0}^{2} = \frac{3\sqrt{3}}{4} (\mu_{0} \gamma_{g})^{2} (M_{0}
H_{c})
\epsilon^{1/2},
\label{om0}
\end{equation}
$Q_{0}$ is the escape point;
\begin{equation}
Q_{0} = \frac{\sqrt{3}}{2} \lambda \epsilon^{1/2} 
\label{cul0}
\end{equation}
and $\lambda$ is again the width of the wall. 
In terms of the microscopic parameters, $B_0 (\epsilon)$ can be rewritten as
\begin{equation}
\frac{1}{\hbar} B_0 (\epsilon)=
\frac{54}{5} \frac{S_{w} \lambda}{\gamma_{g} \hbar}
(M_{0} H_{c})^{1/2} \epsilon^{5/4}  
\sim N_{0} \left( \frac{H_{c}}{M_{0}} \right)^{1/2} \epsilon^{5/4}
\label{215}
\end{equation}
where the second form comes from using the number of spins in the wall, 
$N_{0} = \lambda S_{w}/a^{3}$. 

Although these results were originally derived for the particular case
of a short-ranged defect pinning potential, they obviously 
have more general applicability; as emphasized previously
\cite{stamprev,stampprl}, for small $\epsilon$ almost any pinning potential
acting on the wall via the dipolar field or the wall surface energy will
have the cubic-quadratic form near the coercive field, since its range
will be $\sim \lambda$. However we shall see later in this paper that
the same is {\it not} true of the longitudinal hyperfine field, coming
from the nuclear spins, which fluctuates over much shorter length scales.

Let us at this point introduce 2 examples to which we will return at
various points
in this paper.
We consider two particular systems, Yttrium Iron Garnet (YIG) and
nickel.
YIG is an insulator
with a bcc cubic structure, a saturation magnetisation $\mu_{0} M_{0} =
0.24 \, \mbox{T}$ and with exchange and anisotropy 
energies $J= 1 \times 10^{-11} \, \mbox{J/m}$
and $K_{\|} = 580 \, \mbox{J/m}^{3}$. The width of the domain wall
is $\lambda=860 \AA$, with
a mass per unit area $2 \times 10^{-9} \, \mbox{kg/m}^{2}$.
Nickel is a conductor, again with a cubic structure. The saturation
magnetisation $\mu_{0} M_{0} = 0.6 \, \mbox{T}$, with exchange and anisotropy
$J=3 \times 10^{-11} \, \mbox{J/m}$ and $K_{\|} = 4500 \, \mbox{J/m}^{3}$, 
giving a domain width
$\lambda = 500 \AA$ and a mass $6 \times 10^{-10} \, \mbox{kg/m}^{2}$.
As an example, we give the tunneling rate of
a wall containing $N_{0} = 10^{6}$ spins in nickel. Assuming $\epsilon =
10^{-3}$ and $H_{c}/M_{0} = 0.01$, we get $B_0 /\hbar \sim 20$, and a
frequency $\Omega_{0} \sim 6 \times 10^{9} \, \mbox{sec}^{-1}$ so that
$\Gamma \sim 200 \, \mbox{sec}^{-1}$.

The crossover temperature $T_{0}$ between thermally activated relaxation and
and quantum tunneling is roughly $k_B T_0 \sim \Omega_0 /2 \pi$. 
For many potentials (including the quadratic/cubic potential) there is a fast
crossover between the two modes of relaxation (there 
are however systems,
such as $\mbox{Mn}_{12}\mbox{Ac}$, where tunneling apparently takes
place at intermediary levels, over a wide temperature range). 
For the Ni wall described above, $T_{0} \sim 0.02 \, \mbox{K}$; a
crossover temperature $T_{0} \sim 0.1 \, \mbox{K}$ would require the product
$\mu_{0} H_{c} \epsilon^{1/4} \sim 0.6 \, \mbox{T}$, ie., a coercive 
field similar to the magnetisation density.

\subsubsection{N\'eel Walls in Thin Films}

\begin{figure}
\epsfysize=3.0in
\epsfbox[0 430 510 725]{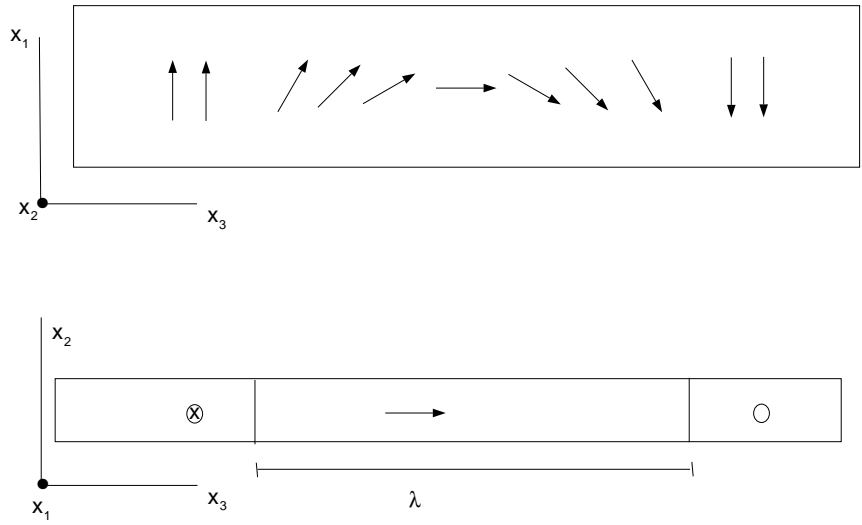}
\caption{The standard model of a N\'eel Wall, with axes labelled as in the
text.}
\label{neel}
\end{figure}

Bloch walls occur predominantly in bulk materials, and so are
difficult to observe individually. Furthermore, they are fairly big, making
the observation of tunneling difficult. If one tries to reduce the size of the
sample to a wire or a platelet, then the Bloch wall becomes unstable and the
magnetisation profile becomes quite complicated, most of the time being
two-dimensional. One alternative is to go to thin films,
of width $\delta \ll \lambda$, and with an anisotropy axis in the plane of the
film. In this case, the strong demagnetisation field forces
the rotation of the magnetisation between
two domains to also take place in the plane of the film, in a head- on fashion.
This is the model
of the N\'eel wall \cite{odell,neel2}. 
Its energy characteristics (ie., the Walker limiting
velocity and the Doring mass), as well as the magnetisation profile  are
similar to that of a Bloch wall.
We consider a film of length
$L \gg \lambda$, width $\delta \ll \lambda$ in the $x_{1}-x_{3}$ plane and
with easy axis in the $x_{1}$ direction. The frame of reference of the
wall is related to the Cartesian frame as
$(x_{1},x_{2},x_{3}) = (y,z,x)$. This is shown in Fig. (\ref{neel}).

%\begin{figure}[t]
%\epsfysize=3.0in
%\epsfbox[0 430 510 725]{neel.ps}
%\caption{Model of a N\'eel Wall}
%\label{neel}
%\end{figure}

The continuum Hamiltonian is
\begin{equation}
{\cal H} = \int d{\bf r} [
J ({\bf \nabla m})^{2} - K_{\|}(\sin^{2} \theta \sin^{2} \phi -1)
+ K_{\bot} \cos^{2} \theta ]
\end{equation}
The profile of the magnetisation is thus
\begin{eqnarray}
\hat{m}^{N}_{1} &=& C \tanh \left( \frac{x_{3}-Q(t)}{\lambda} \right)
\nonumber \\
\hat{m}^{N}_{3} &=& \chi \left( 1 - \frac{\dot{Q}^{2}(t)}{8 c_{0}^{2}}
\right)
\mbox{sech} \left( \frac{x_{3} -Q(t)}{\lambda} \right) 
\label{magcompneel} \\
\hat{m}^{N}_{2} &=& C \frac{\dot{Q}(t)}{2 c_{0}} \mbox{sech} \left(
\frac{x_{3} -Q(t)}{\lambda} \right) \nonumber
\end{eqnarray}

It is useful to ask what experimental set-up could test the theory of 
wall tunneling in this films. Stamp
\cite{stamp1} discussed tunneling of a wall from a defect induced
on the surface of the film (the ends of the wall being held fixed by 
artificially constructed ``gate'' defects - this avoids the problem of
edges effects in the film).

\subsubsection{Bloch and N\'eel Walls in Wires}

The experiments of Giordano et. al.
\cite{giordtun,giordesc,giordres,KHong}
suggest that theoretical
investigations of domain wall tunneling in wires, with diameter 
$d<\lambda$, might be very useful. It should immediately be 
emphasised that the question of the actual configuration of the
magnetisation in such a wire is by no means simple, and suggestions that
the magnetisation profile may be simply modelled as either a Bloch or 
N\'eel wall have been severely criticised by Aharoni \cite{aharoni}. 
This point will
arise again later in the present paper, when we come to consider the 
experiments of Giordano et. al. These experiments
were analysed by the authors using the ``displacement tunneling'' model
of wall tunneling \cite{stamprev,stampprl} just discussed; the
influence of both phonons and magnons on the tunneling was ignored, in
conformity with the results derived by Stamp.

Some rather more exotic tunneling processes have also been suggested for
the cases where a simple Bloch or N\'eel  
wall configuration does occur in wires; these were given the 
names ``chirality tunneling'' and ``Bloch coherence''. 
These possibilities were raised
by Braun and Loss \cite{braun2,braun1} and by 
Takagi and Tatara \cite{tatara1}. Since we will argue later in
this paper that chirality tunneling and
Bloch coherence are highly unlikely 
no matter what the wall looks like, the exact nature of the
magnetisation profile is probably not important 
in what follows (on the other hand, it
clearly {\it is} important in the analysis of Giordano's experiments).

We consider first the possibility of chirality tunneling. For a Bloch
wall, chirality tunneling corresponds to a tunneling of the sense of
rotation of the wall (from clockwise to anti-clockwise, or vice-versa).
The dynamical variable is $\phi(t)$, with tunneling being from 
$\phi = \pm \pi/2$ to $\phi = \mp \pi/2$, with $\theta$ constant,
corresponding to the domain wall magnetisation rotating out of the easy
plane. 
The tunneling splitting is
very small - thus, for a wall in Ni wire, the tunneling splitting
$\Delta_{\chi}$ was estimated to be $0.1 \, \mbox{MHz}$ 
(ie., $5 \mu\mbox{K}$, with a
crossover to tunneling below $T_{0} \sim 0.3 \, \mbox{mK}$. 

For a N\'eel wall the situation is similar, except that the energy
barrier is smaller, and so $\Delta_{\chi}$ has been estimated to be as
high as $80 \, \mbox{mK}$. In both these cases the walls are assumed to be
extremely small, containing perhaps $10^{4}$ spins at most.

Braun and Loss also 
considered band or ``Bloch'' coherent motion of a 1-dimensional
N\'eel wall (which they actually refer to as a Bloch wall). 
In the case of a N\'eel wall tunneling, the Berry phase of
the wall enters in an important way; walls with integer phase have a the
usual Brillouin zone, but walls with $1/2$-integer phase have a halved
Brillouin zone and the Bloch band is split into two bands, with opposite
chirality.
If one ignores all possible sources of decoherence from the environment,
then a very small ($10^{4}$ spins) wall in YIG wire was estimated to
have a Bloch bandwidth of $\Delta_{d} \sim 80 \, \mbox{mK}$. 
It was claimed that
this motion should be observable when $T<T_{c} \sim 80 \, \mbox{mK}$.

\section{Coupling to Phonons and Nuclear Spins }
\resetcounters

As noted in section I, all quantum processes are strongly influenced by
the coupling to the environment. Thus the formulae given in section II
are a rough guide only; to properly determine the tunneling rate (or
whether tunneling occurs at all) one must first include dissipative
effects. This is of course even more true of 
any inelastic quantum diffusive
dynamics, which only exists because of the coupling to the environment.

We now introduce the 2 different environments that will be studied.
We begin with a few general remarks and useful formulae for oscillator
baths and spin baths. We then go on to derive the various couplings
between a domain wall and phonons. This is done for both 1-phonon and
2-phonon couplings. Finally, we derive the coupling 
between a domain wall and the nuclear
spin bath , and show how this yields a slowly fluctuating spatially random
potential acting on the wall. 

\subsection{Oscillator Bath and Spin Bath Environments}

As already remarked in section I, we will be ignoring the effects of
magnons, photons and electrons in what follows; in all cases this is
because their dissipative effects are weak. We are nevertheless left
with 2 important environmental couplings. Before giving a detailed
treatment of these, we make some general remarks and recall some
general formulae that will be relevant in the later technical
discussions.

The phonon bath is representative of the oscillator bath model discussed
by Feynman and Vernon
\cite{vernon}, and Caldeira and Leggett \cite{leggettA}. A subtlety that will
arise here is that the individual oscillators do not necessarily
represent individual phonons: they can also represent pairs of phonons
(an analogous situation was encountered in the discussion of magnon
coupling to domain walls, where the oscillators represented triplets of
magnons \cite{stampprl,stamprev}).

The nuclear spins on the other hand, are a good example of a ``spin
bath'', in which each environmental mode behaves as a 2-level system
(with associated 2-level Hilbert space). In general the spin bath
can behave quite differently from any oscillator bath
\cite{prostamp1,stamp1,prostamp2,prostamp3,prostamp4}.

The interaction between the nuclear spins and the phonons is utterly
negligible. Thus the 2 baths act separately on the domain wall. Their
effects are quite different and depend on the different nature of the 2
baths as follows:

\vspace{5mm}

(i) {\bf Oscillator Baths} : This model assumes 
the environment
can be described as a set of non-interacting harmonic oscillators, with
each environmental mode weakly coupled to the system. 
For $N$ delocalised environmental modes, the
coupling between each oscillator and the system
$\sim O(1/\sqrt{N})$ and the
interaction between oscillators is $\sim O(1/N)$. The  
coupling
between system and environment is {\it linear} in the bath
coordinates. We emphasize that the connection between the oscillator modes
and the modes that might appear in some microscopic model of the system
can be non-trivial and non-linear.
Provided such an oscillator bath exists, one may write a
general
Lagrangian now known as the Caldeira-Leggett Lagrangian \cite{leggettA},
\begin{equation}
L_{CL} =\frac{M}{2}\dot{{\bf Q}}^{2}+V({\bf Q}) +\frac{1}{2} \sum_{{\bf k}}
m_{{\bf k}}
( \dot{{\bf x}}_{{\bf k}}^{2} +
+ \omega_{{\bf k}}^{2} {\bf x_{k}}^{2}) +  
\sum_{{\bf k}} [ F_{{\bf k}} ({\bf Q},\dot{{\bf Q}}) {\bf x_{k}}
+ G_{{\bf k}} ({\bf Q},\dot{{\bf Q}}) \dot{{\bf x}}_{{\bf k}}] + 
\Phi ({\bf Q},\dot{{\bf Q}})
\label{cltotal}
\end{equation}
where the environment is represented by the set $ \{ {\bf x_{k}} \}$ and
$\Phi ({\bf Q},\dot{{\bf Q}})$ is a counterterm used when a 
phenomenological equation of
motion is available; it simply cancels the shift in the potential introduced
when the system is coupled to the oscillators. 

An important simplification occur if 
an expansion of $F_{{\bf k}} ({\bf Q},\dot{{\bf Q}})$ and
$G_{{\bf k}} ({\bf Q},\dot{{\bf Q}})$ to first order in ${\bf Q}$ and 
$\dot{{\bf Q}}$ is possible.
A further set of transformations on the bath \cite{leggettprb} eliminates
the coupling of the bath to $\dot{{\bf Q}}$, and then writing 
$F_{{\bf k}} ({\bf Q}) = C_{{\bf k}} {\bf Q}$,
the Lagrangian is written as 
\begin{equation}
L_{C-L} =L_{0}+L_{B}+ \frac{1}{2} {\bf Q} \sum_{{\bf k}} C_{{\bf k}} {\bf x_{k}} -
\frac{1}{2} {\bf Q}^{2} \sum_{{\bf k}} \frac{ C_{{\bf k}}^{2} }{ m_{{\bf k}}
\omega_{{\bf k}}^{2} }
\end{equation}
where $L_{0}$ is the system Lagrangian and $L_{B}$ 
the oscillator bath Lagrangian. The 
reduced density matrix of the system then defines the effective action
\cite{leggettA,vernon}
$S_{eff}=S_{0}+\Delta S_{eff}$, where 
$S_{0}$ is the system action and 
\cite{leggettA}
\begin{equation}
\Delta S_{eff} = \frac{1}{2} \int_{0}^{1/T} d \tau \int_{0}^{1/T}  d
\tau'
\alpha (\tau-\tau') ({\bf Q}(\tau)-{\bf Q}(\tau'))^{2}
\label{dseff}
\end{equation}
with the kernel
\begin{equation}
\alpha(\tau-\tau') = \frac{1}{2\pi} \int_{0}^{\infty} d\omega J(\omega)
D(\omega, |\tau-\tau'|)
\end{equation}
where the environmental
mode propagator $D(\omega,\tau)$ and spectral function $J(\omega)$ are 
\begin{equation}
D(\omega, \tau) = \mbox{cosech} (\omega /2T)
\cosh (\omega (\frac{1}{2T} - | \tau - \tau' |))
\label{boseprop}
\end{equation}
\begin{equation}
J(\omega)=\frac{\pi}{2} \sum_{{\bf k}} \frac{C_{{\bf k}}^{2}}{m_{{\bf k}}
\omega_{{\bf k}}} \delta (\omega-\omega_{{\bf k}})
\label{jomeg}
\end{equation}
Alternatively, 
one can write 
\begin{equation}
\alpha(\tau-\tau') = \frac{1}{4\pi} T \sum_{n} \int_{0}^{\infty} d\omega
\omega J(\omega)
\frac{e^{i \omega_{n} (\tau-\tau')}}{\omega_{n}^{2}+\omega^{2}}
\label{alphakernel}
\end{equation}
where $\omega_{n}= 2\pi n T$ is the bosonic Matsubara frequency
 \cite{mahan}.

It should be emphasised that these effective Lagrangians, and
the functions $\alpha (\tau )$ and 
$J(\omega)$, refer to a Hilbert space for both system and environment
that has already been truncated to low energies.
High energy environmental
modes, as well as higher energy states of the system itself, are
incorporated into the Lagrangian in the form of renormalised couplings.
Consequently the high-energy form of $J(\omega)$ typically has some
smooth cut-off (for example of the form 
$J(\omega) \sim \exp (-\omega/\omega_{c})$). At low frequency, 
$J(\omega)$ often has power law form. Most dissipation is caused by any
term in $J(\omega)$ of Ohmic linear form, ie., for 
$J(\omega) = \eta \, \omega$ where $\eta$ is the classical friction
coefficient. In the absence of Ohmic dissipation, there will always be
superOhmic terms of the form
$J(\omega) \sim \omega^{s}$, with $s > 1$. Occasionally, one may also
have to worry about a sharp jump in $J(\omega)$ at some finite
frequency, usually imposed by a gap in some set of environmental modes.
However in this case (a) other modes will
always give contributions to $J(\omega)$ for $\omega$ below the gap
energy, and (b) higher-order couplings to the same gapped modes
are typically ungapped. Thus, for example, in the case of magnons
coupled to a Bloch wall \cite{stampprl}, 
even though the magnon spectrum is gapped, one
finds that coupling to pair of magnons gives an ungapped superOhmic 
($J(\omega) \sim \omega^{4}$) term, and coupling to triplets of magnons
gives an ungapped Ohmic term. Note, however, that such
non-linear couplings to multiplets of bosonic excitation are always 
temperature-dependent, ie., $J(\omega) \rightarrow J(\omega,T)$.

In a tunneling situation,
the strength of the coupling to the environment is determined by
the dimensionless parameter $\alpha_{t} = \eta/2 M_{w} \Omega_{0}$
for Ohmic coupling \cite{leggettA}, and we can define 
$\beta_{t} = \tilde{\beta} \Omega_{0}/M_{w}$ in the 
case of superOhmic dissipation with $s=3$, when the 
spectral function $J(\omega) = \tilde{\beta}
\omega^{3}$. 

In the presence of friction, the crossover temperature from quantum
tunneling to thermally activated relaxation is decreased. The new crossover
temperature can be estimated in the case where 
$J(\omega)$ is T-independent,  \cite{weisslivre,hanggi}
\begin{equation}
T_{c} = T_{0} [ (1+\alpha_{t}^{2})^{1/2} - \alpha_{t} ]
\label{tcohmic}
\end{equation}
for Ohmic dissipation. For superOhmic dissipation, in the weak dissipation
regime ($\beta_{t} \ll 1$), $T_{c} = T_{0} (1-\beta_{t}/2)$, 
while for strong coupling $T_{c} = T_{0}/\beta_{t}^{1/3}$ (again for a
T-independent $J(\omega)$). 

\vspace{5mm}

(ii) {\bf Spin Baths}: In a wide variety of cases the environmental
modes each have a finite Hilbert space. In the simplest case where this Hilbert
space is 2-dimensional, each environmental mode is equivalent to some 
2-level system, or to a spin-$1/2$. Examples of this include nuclear
spins, paramagnetic impurities, as well as a variety of more
subtle 2-level systems existing in glasses or disordered solids, often
associated with defects. One should also note the various attempts to
model the behaviour of dissipative quantum systems in terms of a set of
2-level ``Landau-Zener'' degrees of freedom \cite{zenerdiss}. 

Just as for the oscillator bath environment in
Eq. (\ref{cltotal}), one may
write a a general description of a system interacting with a spin bath.
It is more convenient to use a Hamiltonian formulation of this; the most
general Hamiltonian form is then
\begin{equation}
H_{SB} = H_0 ({\bf P},{\bf Q}) +
H_B (\{ \mbox{\boldmath $\sigma$}_{k} \} )
+ H_{int}  ({\bf P},{\bf Q}, \mbox{\boldmath $\sigma$}_{k} \} )
\end{equation}
where ${\bf P}$ is conjugate to the system coordinate ${\bf Q}$, $H_0$
describes the system, and the spin bath has an Hamiltonian
\begin{equation}
H_{B} ( \{ \mbox{\boldmath $\sigma$}_{k} \} ) =
\sum_{k} {\bf h}_{k} \cdot \mbox{\boldmath $\sigma$}_{k} +
\frac{1}{2} \sum_{k} \sum_{k'} V_{k k'}^{\alpha \beta}
\sigma_{k}^{\alpha} \sigma_{k'}^{\beta}
\end{equation}
in which the $\mbox{\boldmath $\sigma$}_{k}$ are Pauli spin matrix
dynamical variables (with $k=1,2,\ldots N$), and the 
${\bf h}_{k}$'s are ``external fields'' which may exist.
The couplings $V_{k k'}^{\alpha \beta}$ between the spins 
are often assumed to
be much weaker than the characteristic energy scales of the system (for
nuclear spins they are the very weak internuclear dipolar interactions;
$| V_{k k'}| \sim 10^{-5} \, \mbox{K}$ or less).

The interaction Hamiltonian has the form
\begin{equation}
H_{int}  ({\bf P},{\bf Q}, \mbox{\boldmath $\sigma$}_{k} \} ) =
 \sum_{k} \left( F_{k}^{z} ({\bf P}, {\bf Q}) \sigma_{k}^{z}
+ \frac{1}{2} ( F_{k}^{+} ({\bf P}, {\bf Q}) \sigma_{k}^{-} +
F_{k}^{-} ({\bf P}, {\bf Q}) \sigma_{k}^{+} ) \right)
\end{equation}
containing both ``diagonal'' couplings
$ F_{k}^{z} ({\bf P}, {\bf Q})$ which polarise the environmental spins)
and ``non-diagonal'' couplings
$F_{k}^{\pm} ({\bf P}, {\bf Q})$ (which flips them).

Whereas in the case of oscillator baths, averaging over the oscillators
is conveniently done using path integrals (to produce an  
``influence functional''), the lack of a well-defined classical path for
the 2-level system makes such an approach infeasible here. Moreover, we
stress that the couplings 
$ F_{k}^{\alpha} ({\bf Q})$ are in general not small. This is in sharp
contrast to the oscillator bath model, where the couplings
$F_{{\bf k}}({\bf Q}) \sim O(N^{-1/2})$; in the spin bath the
couplings are independent of $N$ (so that their effects become more and
more serious as $N$ increases). For hyperfine interactions,
$ F_{k}^{\alpha} ({\bf Q})$ can be as high as $10 \, \mbox{GHz}$ 
($\sim 0.5 \, \mbox{K}$), for 
Ho nuclei - thus one can not always assume that a perturbative treatment
of the $ F_{k}^{\alpha} ({\bf Q})$ is valid. 

To understand the effect of the spin bath on the dynamics of
${\bf Q}$, we make the assumption that
$F_{k}^{z} ({\bf P}, {\bf Q}) \rightarrow
F_{k}^{z} ({\bf Q})$. This is true for all cases studied so far
(including domain walls). In this case, we may distinguish 2 effects.

On the one hand, the diagonal coupling
$U({\bf Q}) = \sum_k F_{k}^{z} ({\bf Q}) \sigma_{k}^{z}$  acts as a
random potential in the coordinate ${\bf Q}$; we shall see this in
detail for the case of a single domain wall. It is useful to define a
density of states $W(U)$ for the potential $U$. In an ensemble of
potentials (or of different systems), and assuming a typical value
$\omega_0$ for $ F_{k}^{z} ({\bf Q})$, $W(U)$ is 
Gaussian in shape, of form
\begin{equation}
W(U) \sim (2 \pi E_{0}^{2} )^{-1/2} \exp [ -U^2 /2 E_{0}^{2} ]
\end{equation}
where $E_0 \sim N^{1/2} \omega_0$ is the gaussian half-width. In general
the potential will fluctuate in time, because of transverse spin
relaxation, at a rate governed by the nuclear 
$T_{2}^{-1}$ (we will ignore $T_1$ processes
in this paper, assuming that $T_1$ is very long).
Thus $U({\bf Q})$ also varies in time,
and fluctuates, for any particular ${\bf Q}$, throughout the domain of
$W(U)$. If $U({\bf Q})$ were static, it would simply block the quantum
dynamics of the system, by removing the degeneracy between initial and
final tunneling states (we call this effect ``degeneracy blocking'').
Its fluctuations in time mean that ``resonance windows'' are
occasionally opened between such states. Notice that the effect of
$U({\bf Q})$ is entirely elastic - no dissipation is involved.

On the other hand the non-diagonal couplings 
$ F_{k}^{\pm} ({\bf P},{\bf Q})$ cause 
environmental spins to {\it flip} under the action of the system. 
Thus, whereas the main effect of $F_{k}^{z}({\bf Q})$ is to completely alter
the effective potential acting on the system, the main effect of the
non-diagonal couplings is to cause transitions in the spin bath, leading
to dissipation and/or decoherence. Unlike the oscillator bath,
dissipation and decoherence are not necessarily linked in the spin bath
- indeed it is possible in certain cases to have decoherence without
dissipation. Because the main effect of transitions in the spin bath is
the absorption of a random topological Berry phase by the bath
(thereby randomising the quantum phase associated with motion in 
${\bf Q}$),
this decoherence is called ``topological decoherence''.

In this
paper we shall develop the theory for a domain wall coupled to nuclear
spins, but we  shall ignore the dissipative effects coming from nuclear
transitions. While these are not necessarily small, the ``degeneracy
blocking'' effects, coming from the diagonal coupling 
$F_{k}^{z}({\bf Q})$ are often considerably larger, particularly for 
weak hyperfine potentials (such as in Ni or Fe materials). Their main
effect is the generation of a random 
potential acting on the wall, whose effects must be understood before
dissipation is considered. 

\subsection{Phonon Couplings to the Wall}

As noted above, it is not sufficient to consider only 1-phonon couplings
to the wall; we must also consider non-linear 2-phonon couplings. These
will be analysed using standard continuum elasticity theory for acoustic
phonons coupled to the magnetisation; this rather intricate subject is
reviewed by de Lacheisserie \cite{tremolet1}

The Euclidean Lagrangian of the acoustic phonons in a material of mass
density
$\rho_{v}$ is 
\begin{equation}
{\cal L}_{p}({\bf x},\tau) = \frac{1}{2} \rho_{v} \dot{u}_{i}^{2} +
\frac{1}{2}
C_{ijkl} U_{ij}U_{kl}
\end{equation}
where $u_{i}={\bf r}_{i}-{\bf r}_{i}^{0}$ represents the phonon's field, ie.,
the displacement of the atoms with respect to their equilibrium position, and
$U_{kl} = (\partial_{k}u_{l} + \partial_{l}u_{k})/2$ is the strain tensor.
The potential energy of the field is given by the elastic tensor
$C_{ijkl}$, with typical values $C \sim 10^{11} \mbox{J/m}^{3}$. 
The indices refer to the directions of the displacements, with
the summation convention over repeated indices applied.
We begin by considering an isotropic elastic energy, which is the simplest
form
\begin{equation}
C_{ijkl}=\lambda_{e} \delta_{ij} \delta_{kl} +
\mu_{e} ( \delta_{ik}\delta_{jl} + \delta_{il} \delta_{jk})
\end{equation}
where $\mu_{e}$ and $\lambda_{e}$ are the Lam\'{e} constants.
The tensors that are going to be considered in this paper are all
symmetric under the exchange of two indices forming a pair
$ \{ i j \}$
and it is convenient to use the abbreviated notation
\begin{equation}
\begin{array}{ccc}
\{ 1 1 \} =1 \; & \{ 2 2 \} =2 \; & \{ 3 3 \} =3 \\
             &              &              \\
\{ 2 3 \} =4 \; & \{ 1 3 \} =5 \; & \{ 1 2 \} =6
\end{array}
\label{indices}
\end{equation}
We will thus represent any pair of indices by a letter
$a = \{ a_{1} a_{2} \} = \{ a_{2} a_{1} \} = 1, ..., 6$. This simplifies
the notation tremendously.

We can also consider the phonons of a
cubic structure. In this case, the non-zero
elastic constants are $C_{11}=C_{22}=C_{33}$,
$C_{12}=C_{13}=C_{23}= \ldots$ and
$C_{44}=C_{55}=C_{66}$. It reduces to the
isotropic case if $C_{11}-C_{12}-2C_{44}=0$, in which case
$C_{12}=\lambda_{e}$ and $C_{44}=2\mu_{e}$.

It is observed experimentally that the shape of a fully magnetised
ferromagnetic substance is different from the shape in the
unmagnetised state. This phenomenon, called magnetostriction, can
be related to the strain dependence of the anisotropy energy.
There is thus an interaction between the direction of the magnetisation and
the strain tensor of a solid. In many cases, this interaction is linear in
the strain tensor. The simplest type of interaction that respects
time-reversal symmetry is thus of the form 
$U_{ij} \hat{m}_{k} \hat{m}_{l}$, with $i,j,k$ and $l$ 
arbitrary directions.  The interaction between phonons and the magnetisation 
is mediated by the first order magnetoelastic tensor, $A_{ijkl}$,
as discussed in de Lacheisserie \cite{tremolet1}.
It is also possible to consider an interaction that is quadratic in the
strain and in the magnetisation, which is described by the
second order magnetoelastic tensor $R_{ijklmn}$ first introduced 
by Mason
\cite{mason} and also discussed by de Lacheisserie \cite{tremolet1}.
The general form of the interaction between the phonons 
and the domain wall is given by integrating over these local tensor
couplings, throughout the sample, for a given wall profile, and then
subtracting off the result that would have been obtained in the same
integration in the absence of the wall. Thus, if the ``vacuum''
magnetisation profile is written as 
${\bf m}^{0}$, the general form of the 
interaction Lagrangian will be
\begin{equation}
L_{int} = - \int d^{3} {\bf r} 
A_{ijkl} U_{ij} ({\bf r}) 
(\hat{m}_{k} ({\bf r}) \hat{m}_{l} ({\bf r}) -
\hat{m}_{k}^{0} \hat{m}_{l}^{0})
+ R_{ijklmn} U_{ij} ({\bf r}) U_{kl} ({\bf r}) 
(\hat{m}_{m} ({\bf r}) \hat{m}_{n} ({\bf r}) -
\hat{m}_{m}^{0} \hat{m}_{n}^{0})
\label{mgel}
\end{equation}
up to 2nd order in the phonon strain variables. The ``vacuum'' (no wall)
term is just a constant (the integrated bulk magnetoelastic stress
energy for the entire sample), which we subtract off in all our
calculations. 

>From now on, we will use the definitions of the
indices in Eq. (\ref{indices}) to discuss this Lagrangian.
The first order tensor has been extensively studied in the context of
magnetostriction. 
The second order tensor is encountered less frequently, but has been used
in connection with the so-called ``morphic effect'', ie., the change in 
sound velocity as a function of the direction of the 
applied field \cite{mason}.
Both of these couplings come from an expansion
of the exchange energy in terms of the displacement of the atoms. In our
context, $A$ and $R$ then correspond to 1- and 2-phonon interaction terms
respectively. Note that 2-phonon means that the interaction involves
two phonons simultaneously, a process quite different from two 1-phonon
processes \cite{KagProk,proko2}.

For a crystal with orthorhombic
symmetry, there are 12 non-zero matrix elements: $A_{11}$, $A_{22}$,
$A_{33}$, $A_{12}$, $A_{21}$, $A_{13}$, $A_{31}$, $A_{23}$, $A_{32}$,
$A_{66}$, $A_{55}$ and $A_{44}$. If the symmetry of the crystal is
reduced to
cubic, then $A_{11}=A_{22}=A_{33}$, $A_{44}=A_{55}=A_{66}$ and
$A_{12}=A_{21}=A_{31}...$. The second order magnetoelastic tensor is
obviously
very complicated, but we will only use $R_{111} = R_{222} = R_{333}$ with all
other $R_{abc}$ equal to zero. This
approximation is justified by the fact that these three components are
generally
of the same order of magnitude, while being at least two orders of magnitude
larger than the other coefficients.
Several measurements and calculations of
these coefficients have been performed \cite{neel1,callencallen}.
In YIG, these are \cite{eastman}: $A_{ab} \sim 10^5 \, \mbox{J/m}^{3}$,
$R_{111} \sim  10^{7} \, \mbox{J/m}^{3}$, with
the transverse sound velocity \cite{clark}: $c_{T} \sim 3 \times 10^{3}
\, \mbox{m/sec}$. In nickel \cite{tremolet2}:  $A_{ab} \sim 10^{8} \, 
\mbox{J/m}^{3}$,
$R_{111} \sim 10^{10} \, \mbox{J/m}^{3}$, and $c_{T} \sim 10^{3} \, 
\mbox{m/sec}$. 

We use the Fourier transform of the phonon field
\begin{equation}
{\bf u}({\bf r}, \tau) = T \sum_{n} \sum_{{\bf q}}
e^{-i\omega_{n} \tau + i {\bf q} \cdot {\bf r}} \;
{\bf u}({\bf q}, i\omega_{n})
\end{equation}
to write the action corresponding to the interaction
Lagrangian, Eq. (\ref{mgel}) as
$S_{I}[{\bf u},\hat{\bf m}]= S_{I}^{(i)}[{\bf u},\hat{\bf m}] +
S_{I}^{(ii)}[{\bf u},\hat{\bf m}]$ with
\begin{equation}
S_{I}^{(i)}[{\bf u},\hat{{\bf m}}] = \frac{i}{2} A_{ab} T
\sum_{n} \sum_{{\bf q}} \int_{0}^{1/T}
d\tau e^{i {\bf q} \cdot {\bf Q}(\tau)} e^{-i \omega_{n} \tau}
{\cal M}_{b}(-{\bf q})
\left[ q_{a_{1}}u_{a_{2}}({\bf q},i\omega_{n}) +
q_{a_{2}}u_{a_{1}}({\bf q},i\omega_{n}) \right]
\label{smgel1}
\end{equation}
representing the action coming from 1-phonon processes and
\begin{eqnarray}
& & S_{I}^{(ii)}[{\bf u},\hat{{\bf m}}] = -\frac{1}{4} R_{abc} T^{2}
\sum_{n n'} \sum_{{\bf k}{\bf k}'}
\int_{0}^{1/T} d\tau e^{i ({\bf k}+{\bf k}') \cdot {\bf Q}(\tau)}
e^{-i (\omega_{n} +\omega_{n'}) \tau}
{\cal M}_{c}(-{\bf k}-{\bf k}') \times \nonumber \\
& &
\left[ k_{a_{1}}u_{a_{2}}({\bf k},i\omega_{n}) +
k_{a_{2}}u_{a_{1}}({\bf k},i\omega_{n}) \right]
[ k_{b_{1}}'u_{b_{2}}({\bf k}',i\omega_{n}') +
k_{a_{2}}'u_{b_{1}}({\bf k}',i\omega_{n}') ]
\label{smgel2}
\end{eqnarray}
corresponding to 2-phonon processes. ${\bf Q}(\tau)$ is the position of the wall
and the summation convention over repeated magnetoelastic indices
$a=\{ a_{1} a_{2} \}$ is in effect.

In these two expressions, the profile of the magnetisation is included
in the magnetisation form factor ${\cal M}_{a}$ defined as
\begin{equation}
{\cal M}_{a}({\bf q}) = \int d^{3}{\bf r} e^{-i {\bf q} \cdot {\bf r}}
( \hat{m}_{a_{1}}({\bf r}) \hat{m}_{a_{2}}({\bf r}) - 
  \hat{m}_{a_{1}}^{0} ({\bf r}) \hat{m}_{a_{2}}^{0}({\bf r}))
\end{equation}

The magnetoelasticity of thin films differs considerably from what is observed
in the bulk \cite{tremolet1}.
The elastic properties and the symmetry of a film will in general be different
than those in bulk, and the strain associated with the surface will also play a
major role. The magnetostriction constants in general are dependent on the
thickness of the film and, furthermore, in many cases when discussing the
simple magnetostriction, it is impossible to consider only the linear
magnetostriction.
It turns out that it is strain dependent and that the second order
magnetoelastic constants play a major role \cite{tremolet1,ohanley}.
It is obvious that a proper discussion of the magnetoelastic dissipation
requires a very detailed knowledge of the material being used. To keep the
discussion general, we still use the Lagrangian  Eq. (\ref{mgel}) where
now the phonons will be 2-dimensional, with a surface density
$\rho_{s}$ and the elastic and magnetoelastic constants
have units of $\mbox{J/m}^{2}$. Similarly, in 1 dimension, we consider a
linear density $\rho_{l}$ and the constants have units of $\mbox{J/m}$.
More detailed models could be devised if necessary.

Finally, a special comment must be made about the terms in 
the magnetisation profile
appearing due to the finite velocity of the wall, ie., coming
from the demagnetisation energy. They represent a coupling to the
environment that is proportional to the velocity of the macroscopic
coordinate. The standard way to deal with this coupling is to
introduce a total time derivative term in the action and to perform a
canonical transformation to a new set of oscillators (c.f. Leggett
\cite{leggettprb}).
In our case, it does not appear that this can be done easily since the
coupling is not ``strictly linear'', but is a complicated function of
${\bf Q}$.
We will first build the effective action of the magnetisation including 
the velocity term.  This will allow us to see in which cases 
exactly this coupling to the velocity is important. Once its physical 
relevance or irrelevance is established, it is easy to go back to 
the original Lagrangian and to deal with this term directly.

\subsection{Nuclear Spin Coupling to the Wall} 

Nuclear hyperfine effects vary enormously between magnetic systems.
The weakest is in Ni, where only $ 1 \%$ of the nuclei have spins,
and the hyperfine coupling is only 
$\omega_{0} = 28.35 \, \mbox{MHz}$ 
($\sim 1.4 \, \mbox{mK}$). On the other hand, in the
case of rare earths, $\omega_{0}$ varies from $1$ to $10 \, \mbox{GHz}$ 
($0.05$ to $0.5 \, \mbox{K}$). 
In this latter case the hyperfine coupling energy to a {\it single}
nucleus may be comparable to the other energy scales in the problem !

To derive an effective interaction Hamiltonian, we consider our system
of ferromagnetically ordered spins to coupled locally to $N$ nuclear
spins 
${\bf I}_k$ 
at positions ${\bf r}_{k}$ ($ k=1,2,3, ... \, N$), which for a set
of dilute nuclear spins (where only one isotope has a nuclear spin) will
be random. The total Hamiltonian for the coupled system is then
\begin{equation}
H = H_{m} + \sum_{k=1}^{N} \omega_{k} {\bf s}_{k} \cdot 
{\bf I}_{k} + 
\frac{1}{2} \sum_{k} \sum_{k'} V_{k k'}^{\alpha \beta}
I_{k}^{\alpha} I_{k'}^{\beta}
\label{nusp}
\end{equation}
where $H_{m}$ is the electronic Hamiltonian for the magnetisation 
(Eq. (\ref{hamiltonian})), written in terms of the electronic spins
${\bf s}_k$ at the sites where there happen to be nuclear
spins,
and $\omega_{k}$ is the hyperfine coupling at
${\bf r}_{k}$; $V_{k k'}^{\alpha \beta}$ is the internuclear dipolar
interaction,, with strength 
$|V_{k k'}^{\alpha \beta}| \sim 1-100 \, \mbox{kHz}$ ($0.05-0.5 
\, \mu\mbox{K}$). 
In terms of the continuum magnetisation
${\bf M}({\bf r})$, we have
\begin{equation}
H=H_{m} + \sum_{k=1}^{N} \omega_{k} 
\int \frac{d^{3}r}{\gamma_{g}} \delta ({\bf r}-{\bf r}_{k}) 
[ M_{z}({\bf r}) I_{k}^{z} + ( M_{x}({\bf r}) I_{k}^{x} 
+  M_{y}({\bf r}) I_{k}^{y})] +
\frac{1}{2} \sum_{k} \sum_{k'} V_{k k'}^{\alpha \beta}
I_{k}^{\alpha} I_{k'}^{\beta}
\label{inutile}
\end{equation}

The resemblance between this form and the spin bath coupling form given
above is evident - however, it is often more convenient to rewrite 
Eq. (\ref{inutile}) by separating out the slowly-varying wall profile
${\bf M}_{W}({\bf r})$ from the ``fast'' magnon fluctuations which ride
on top of this. Thus we write
\begin{equation}
{\bf M}({\bf r}) = {\bf M}_{W}({\bf r},Q) + 
\mbox{\boldmath $\delta${\bf M}}({\bf r}) 
\end{equation}
where ${\bf M}_{W}({\bf r},Q)$ is the wall profile for a wall centered at
$Q$ (given in Eq. (\ref{magcomp})  
for a Bloch wall and Eq. (\ref{magcompneel}) for a N\'eel wall). We then
introduce local axes $(u,v,w)$, with $\hat{{\bf x}}_{w}$ parallel to 
${\bf M}_{W}({\bf r},Q)$, and $\hat{{\bf x}}_{u}$, $\hat{{\bf x}}_{v}$
perpendicular to it, and to each other (these axes were called 
$x_{1},x_{2},x_{3}$ in previous papers \cite{stamprev,stampprl}). 
Then we quantise the 
$\mbox{\boldmath $\delta$M}({\bf r})$ in terms of local magnon
operators $b ({\bf r})$ and $b^{+} ({\bf r})$, as usual:
\begin{eqnarray}
\delta M_{w}({\bf r}) &=& -4 \gamma_{g} 
b^{+} ({\bf r}) b ({\bf r}) \nonumber \\
\delta M_{+}({\bf r}) &=& (4 \gamma_{g} M_{0})^{1/2}
\left(  1 - \frac{2 \gamma_{g}}{M_{0}} b^{+} ({\bf r}) b ({\bf r})
\right)^{1/2} b ({\bf r}) \\
\delta M_{-}({\bf r}) &=& (4 \gamma_{g} M_{0})^{1/2}
b^{+} ({\bf r}) 
\left(  1 - \frac{2 \gamma_{g}}{M_{0}} b^{+} ({\bf r}) b ({\bf r})
\right)^{1/2} \nonumber 
\end{eqnarray}
where $\delta M_{\pm}({\bf r}) = \delta M_{u}({\bf r}) \pm 
\delta M_{v}({\bf r})$.
The interaction term in Eq. (\ref{inutile}) is now written for a single
domain wall by again subtracting off the ``vacuum'' magnetisation
profile ${\bf M}_{0}$ obtaining 
in the absence of the wall, to get a sum of longitudinal and 
transverse couplings:
\begin{equation}
H_{int}(Q,\{ {\bf I}_{k} \}) = ^{\|} \hspace{-0.1cm}H_{int}+^{\bot}H_{int}
\end{equation}
\begin{equation}
^{\|}H_{int} = \sum_{k=1}^{N} \omega_{k} 
\int \frac{d^{3}r}{\gamma_{g}} \delta ({\bf r}-{\bf r}_{k})
\left( (M_{W}({\bf r},Q) I_{k}^{w} - {\bf M}_{0} \cdot 
{\bf I}_{k} ) + I_{k}^{w} \delta M_{w}({\bf r})
\right)
\end{equation}
\begin{equation}
^{\bot}H_{int} = \frac{1}{2} \sum_{k=1}^{N} \omega_{k} 
\int \frac{d^{3}r}{\gamma_{g}} \delta ({\bf r}-{\bf r}_{k}) 
\left( \delta M_{+}({\bf r}) I_{k}^{-} + 
\delta M_{-}({\bf r}) I_{k}^{+}
\right)
\end{equation}

To get an idea of the size and effect of each of these terms, let us
choose as example a Bloch wall, supposing for simplicity that 
$\omega_{k}=\omega_{0}$ for each nuclear spins (in reality the 
$\{ \omega_{k} \}$ will be spread out by the internuclear 
$V_{kk'}^{\alpha \beta}$, as well as by transfer hyperfine couplings to
non-magnetic ions).

Supposing also that the Bloch wall is static, we have, 
from Eq. (\ref{magcomp}), for
a wall centered at $x_{3}=Q$, a static longitudinal interaction
\begin{equation}
^{\|}H_{int}^{Bloch}(Q,\{ {\bf I}_{k} \}) = 
\frac{\omega_{0} M_{0}}{\gamma_{g}} \sum_{k=1}^{N}
\int d^{3}r \delta ({\bf r}-{\bf r}_{k}) 
\left( \left( 1 - C \tanh \left(\frac{x_{3}-Q}{\lambda_{B}}\right)
\right) + \delta m_{w}({\bf r}) \right) I_{k}^{w}
\label{nucspinlong}
\end{equation}
and a transverse interaction
\begin{eqnarray}
^{\bot}H_{int}^{Bloch}(Q,\{ {\bf I}_{k} \}) & = & 
\frac{\omega_{0} M_{0}}{2 \gamma_{g}} \sum_{k=1}^{N}
\int d^{3}r \delta ({\bf r}-{\bf r}_{k}) 
\left( \chi \mbox{sech}  \left(\frac{x_{3}-Q}{\lambda_{B}}\right)
 \left( I_{k}^{+} + I_{k}^{-} \right) \right. \nonumber \\
& & + \left. \left( \delta m_{-}({\bf r}) I_{k}^{+} +
\delta m_{+}({\bf r}) I_{k}^{-} \right)
\frac{}{} \right)
\label{nucspintrans}
\end{eqnarray}
where $\mbox{\boldmath $\delta${\bf m}} = 
\mbox{\boldmath $\delta${\bf M}}/M_{0}$.

We can extract from Eq. (\ref{nucspinlong}) and 
(\ref{nucspintrans}) the following 3 terms:

(i) Terms of the form 
$I_{k}^{\alpha} \mbox{\boldmath $\delta${\bf m}}_{\beta}$ couple nuclear 
spins
to magnons; this interaction is considerably enhanced inside the wall
where the magnons are gapless, and is known to be important in the 
discussion of the NMR relaxation rate inside the wall. 

(ii) We also have a longitudinal 
$\tanh[(x_{3}-Q)/\lambda_{B}]$ term in $^{\|}H_{int}$ which is diagonal
in the local nuclear spin basis states (with axis of quantisation 
$\hat{\bf x}_{w}$ along ${\bf M}_{W}({\bf r},Q)$). This term gives us a
random potential when summed over all spins, which we write as 
\begin{equation}
U(Q) = \frac{ \omega_{0} M_{0}}{\gamma_{g}} \sum_{k=1}^{N} \int
d^{3}r \delta ({\bf r}-{\bf r}_{k}) 
\left( 1 - C \tanh \left(\frac{x_{3}-Q}{\lambda_{B}}\right)
\right)I_{k}^{w}
\label{pinpotnucspin}
\end{equation}

\begin{figure}[ht]
\epsfxsize=6.5in
\epsfbox[-110 150 593 653]{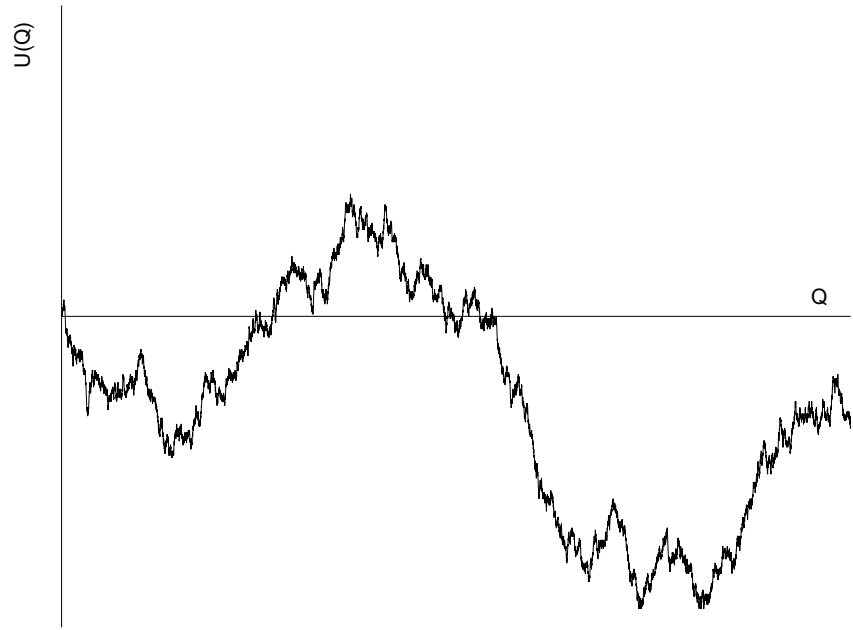}
\caption{Random potential caused by the nuclear spins when
$kT \gg \omega_0$. At any given time the potential has a "random
walk" form (the variance in 
spatial fluctuations of the potential increase as
the square root of the distance travelled by the wall). The potential also
slowly fluctuates in time, at a rate determined by nuclear spin
diffusion.}
\label{random}
\end{figure}

Suppose we use the ``high-T'' limit, where $kT \gg \omega_{0}$, so that
the expectation value $\langle {\bf I}_{k} \rangle$ is zero in any
direction. Since the nuclear spins are completely uncorrelated 
($V_{kk'}^{\alpha \beta}$ is negligible), a volume containing $N$
nuclear spins will have root mean square polarisation $\sim N^{1/2}$. It
then immediately follows that for a system in which a fraction $x$ of
all the states are occupied by nuclear spins ${\bf I}_{k}$ at sites $k$,
the ensemble averaged correlation between $U(Q)$ for 2 different
wall position is 
\begin{eqnarray}
{\cal C}_{UU}(Q_{1}-Q_{2}) = 
\langle \left( U(Q_{1})-U(Q_{2}) \right)^{2} \rangle &\sim&
\omega_{0}^{2} s^{2} I^{2} \Delta N_{Q_{1}-Q_{2}} \nonumber \\
&\sim& \left( \frac{\omega_{0} M_{0} I}{\gamma_{g}} \right)^{2}
a_{0}^{3} x S_{w} \left( Q_{1}-Q_{2} \right) \hspace{5mm} 
(T \gg \omega_0 )
\label{correlation}
\end{eqnarray}
where $s = | {\bf s}_{k} |$, $I = | {\bf I}_{k} |$, $S_{w}$ is the area
of the wall, and 
$\Delta N (Q_{1}-Q_{2}) \sim (x S_{w}/a_{0}^{3})(Q_{1}-Q_{2})$ is the
mean number of nuclear spins in the volume swept out by the wall in
going from $Q_{1}$ to $Q_{2}$. Thus $U(Q)$ exhibits ``random walk''
behaviour over the sample \cite{Bouchaud}, as represented on Fig. 
(\ref{random}). 

%\begin{figure}[ht]
%\epsfxsize=6.5in
%\epsfbox[-110 150 593 653]{random.ps}
%\caption{Random potential caused by the nuclear spins when 
%$kT \gg \omega_0$. The mean fluctuations of the potential increase as 
%the square root of the distance travelled by the wall.}
%\label{random}
%\end{figure}

Eq. (\ref{correlation}) is for the high-T limit.
In the case of very low T, when
$T \ll \omega_0$, the nuclear spins will tend to line up in the
field due to the electronic moments. In the low-T limit, they will all
be aligned according to the wall configuration, and Eq.
(\ref{correlation}) is replaced by
\begin{eqnarray}
{\cal C}_{UU}(Q_{1}-Q_{2}) \sim
\left( \frac{\omega_0 M_0 I}{\gamma_g} \right)^{2}
a_{0}^{6} x^2 S_{w}^{2} (Q_{1}-Q_{2})^{2}
\label{correlationlow}
\end{eqnarray}
Thus the wall is trapped in a potential which increases linearly (on
average) in both directions away from the wall centre. In reality, both
results  (\ref{correlation}) and  (\ref{correlationlow}) are only valid
for times $\ll T_2$; for larger
times the internuclear dipolar transitions cause random  
fluctuations of the hyperfine potential.

(iii) Finally, we have the $\mbox{sech} [(x_{3}-Q)/\lambda_{B}]$ 
term in 
Eq. (\ref{nucspintrans}), acting only inside the wall; this simply tells
us that in the chirally-rotated frame of ${\bf M}_{W}({\bf r},Q)$, the
nuclear spins quantised along the ``vacuum'' ${\bf M}_{0}$ are no longer
diagonal. Because of this term, a domain moving through a field of
nuclear spins can flip some of them as it passes, thereby causing both
topological decoherence and dissipation in general. From the solution of
the ``central spin'' problem \cite{prostamp1}
one may estimate that the mean number
$\lambda_{I}$ of flipped spins caused by a single excursion between 2
points $Q_{1}$ and $Q_{2}$ (sweeping out $\Delta N$ nuclear spins) will
be given by 
\begin{equation}
\lambda_{I} \sim \frac{1}{2} \Delta N 
\left( \frac{\pi \omega_{0}}{2 \Omega_{0}} \right)^{2}
\label{linumb}
\end{equation}
where $\Omega_{0}$ is the typical frequency involved in the motion from
$Q_{1}$ to $Q_{2}$ ($\Omega_{0}$ being the bounce frequency for tunneling
motion), provided $\omega_{0} \ll \Omega_{0}$.

To get a feeling for the numbers, 
consider the ``high-$T$'' limit $k_B T \gg \omega_{0}$.
Now imagine a wall of area 
$(300 \times 300) \AA^{2}$, which sweeps out a length
$Q_{1}-Q_{2} \sim 300 \AA$, and a volume containing $\sim 10^{6}$
electronic spins (ie., $a_{0} \sim 3 \AA$); suppose also that 
$S = | {\bf S}_{k} | =1 $ and $I = | {\bf I}_{k} |=1/2$. We may imagine 2
extreme cases: 

(a) Suppose $\omega_{0}$ is very small, eg., let $\omega_{0} =1.4 \, 
\mbox{mK}$, and let $x=0.01$ (only $1 \%$ of nuclei have spins); these
are values appropriate to Ni. Then the Gaussian mean change in $U(Q)$
for the wall will be $|\Delta U| \sim 0.3 \, \mbox{K}$, in the high-$T$
limit (ie., $T \gg 1.4 mK$; all experiments up to now have been in this
limit).
If we assume
that $\Omega_{0} \sim 2 \times 10^{10} \, \mbox{Hz}$ (typical of wall
tunneling), then $\lambda_{I} \sim 0.025$, ie., in most cases no nuclei
will be flipped at all.

(b) Suppose now $\omega_{0} = 0.2 \, \mbox{K}$, a value more typical of
some rare earth magnets, and that $x=1$. We than have,
in the high-$T$ limit (now for $T \gg 0.2K$), that 
$|\Delta U| \sim 400 \, \mbox{K}$, and $\lambda_{I} \sim 10^{5}$ (each
nucleus has a probability $\sim 0.1$ of being flipped) ! Even if we
reduce the displacement in this case from $300 \AA$ to only $3 \AA$
(ie., the wall makes only a single lattice displacement), we still find
$|\Delta U| \sim 40 \, \mbox{K}$ and $\lambda_{I} \sim 10^{3}$.

Notice furthermore that the low-$T$ limit $k_B T \ll \omega_{0}$, 
where all nuclear spins line
up with the local magnetisation, is easily attainable for rare
earth systems, in current
experiments. In this case the effects are spectacular; for the present
example, with $\omega_{0} = 0.2 \, \mbox{K}$, a displacement of $300 \AA$
leads to an energy change $|\Delta U| \sim 20 \, \mbox{eV}$!! Thus the
wall is completely locked to the nuclear polarisation when 
$k_B T \ll 0.2 \, \mbox{K}$.

Thus in the first case, of extremely weak coupling, we see that dynamic
effects in the bath can be ignored - the sole function of the nuclear
bath is to provide a weak random static potential. On the other hand in
the case of rare-earth magnets, with very strong hyperfine coupling, the
nuclear effects are quite enormous, and moreover it is crucial to take
account of nuclear transitions in analysing the wall motion. 

In this paper we will concentrate on the weak coupling case, because it
is simpler and more relevant to the Ni wire experiments of
Giordano et. al. We shall deal in section V with the
residual effects of the internuclear coupling, which drives the slow (on
the scale of $\Omega_0^{-1}$) fluctuations in time of $U(Q)$.

\section{Effective Action of a Domain Wall coupled to Phonons}

In this section, we obtain the effective action of a domain wall interacting
with the phonons through the magnetoelastic interaction \cite{tremolet1};
we completely ignore the nuclear spins in this section. 
We consider both
1-phonon and 2-phonon processes, as described in the last section, and 
we relate
their effect to the phenomenological oscillator bath of Caldeira and 
Leggett \cite{leggettA}. As expected 1-phonon processes give rise to 
superOhmic dissipation with $J(\omega) \sim \omega^{3}$. We find 
however that the dominant contribution to the dissipation comes from the
coupling between the {\em velocity} of the wall and the phonons. 

The second order magnetoelastic tensor brings in
2-phonon mediated Ohmic dissipation, with a
temperature dependent friction coefficient $\eta(T) \sim T^{3+d}$ 
(where $d$ is the effective 
dimensionality of the phonons), just as with the quantum diffusion of a 
particle in an insulator \cite{proko2}. In this case however, 
the coupling to the 
velocity of the wall is unimportant. 
As part of our the discussion of 
2-dimensional 2-phonon processes, we discuss the general case of the 
dissipation of a N\'eel wall. This represents a straightforward extension of 
the analysis of a Bloch wall. The same is true for the dissipation in a
wire, where the phonons are effectively 1-dimensional. 

\subsection{Effective Action of a Bloch Wall}

We are interested in integrating out the phonons so as to obtain an 
effective action for 
the magnetisation. 
The effective action is defined as
\begin{eqnarray}
e^{-S_{eff}[\hat{{\bf m}}]} &=& e^{-S_{0}[\hat{{\bf m}}]}
\frac{ \int D[{\bf u}] e^{-S_{0}[{\bf u}]-S_{I}[\hat{{\bf m}},{\bf u}]}}
{ \int D[{\bf u}] e^{-S_{0}[{\bf u}]}} \nonumber \\
&=& e^{-S_{0}[\hat{{\bf m}}]} \langle e^{-S_{I}[\hat{{\bf m}},{\bf u}]} \rangle
\end{eqnarray}
where $S_{0}$ is the action of the phonons and $S_{I}$ is the magnetoelastic 
interaction , as 
described in the Introduction. We recall that
the interaction action $S_{I}=S_{I}^{(i)}+S_{I}^{(ii)}$, representing 1- and
2-phonon processes, is expressed as the sum of Eq. (\ref{smgel1}) and
Eq. (\ref{smgel2}). 
The uniform magnetisation, in absence of a domain wall, is in the
$x_{1}$ direction. 
The form factor of the magnetisation, appearing
in the action $S_{I}$ is thus
\begin{equation}
{\cal M}_{a}({\bf q}) = \int d^{3}{\bf r} e^{-i {\bf q} \cdot {\bf r}} 
(\hat{m}_{a_{1}}({\bf x}) \hat{m}_{a_{2}}({\bf r}) - \delta_{a_{1},1}
\delta_{a_{2},1})
= (2 \pi )^{2} \delta(q_{1}) \delta(q_{2}) {\cal M}_{a}(q_{3})
\label{64}
\end{equation}
This last form coming from the one-dimensional properties of a Bloch wall.
The magnetisation 
depends only on the coordinates normal to the wall and this restricts
the momentum in the plane of the wall to be zero. Using the description of a
Bloch wall in term of the component of the
magnetisation Eq. (\ref{magcomp}), 
it is easy to compute these different terms. Up to second
order in the velocity of the wall, they are: 
\begin{equation}
{\cal M}^{B}_{1}(q_{3})= - \pi q_{3} \lambda^{2}
\mbox{cosech} ( \pi q_{3} \lambda /2)
\label{mdebut}
\end{equation}
\begin{equation}
{\cal M}^{B}_{2}(q_{3},\tau )= \left( 1 - \frac{\dot{Q}^{2}(\tau)}{4c_{0}^{2}} 
\right) \pi q_{3} \lambda^{2} \mbox{cosech} ( \pi q_{3} \lambda /2) 
\end{equation}
\begin{equation} 
{\cal M}^{B}_{3}(q_{3},\tau )= \frac{\dot{Q}^{2}(\tau)}{4 c_{0}^{2}} 
\pi q_{3} \lambda^{2} \mbox{cosech} ( \pi q_{3} \lambda /2)
\end{equation} 
\begin{equation}
{\cal M}^{B}_{4}(q_{3},\tau )= \frac{\dot{Q}(\tau)}{2 c_{0}}
\pi q_{3} \lambda^{2} \mbox{cosech} ( \pi q_{3} \lambda /2)
\end{equation}
\begin{equation}
{\cal M}^{B}_{5}(q_{3},\tau )= -i \frac{\dot{Q}(\tau)}{2 c_{0}}
\pi q_{3} \lambda^{2} \mbox{sech} ( \pi q_{3} \lambda /2)
\end{equation}
\begin{equation}
{\cal M}^{B}_{6}(q_{3},\tau )= -i \left( 1 - \frac{\dot{Q}^{2}(\tau)}
{8 c_{0}^{2}} \right) \pi q_{3} \lambda^{2} \mbox{sech} ( \pi q_{3} \lambda /2)
\label{mfin}
\end{equation}
where $\lambda = (J/K_{\|})^{1/2}$ is the domain wall width and 
$c_{0}$ is the Walker velocity, defined in Eq. (\ref{vwalker}). 
There are essentially two types of coupling; those proportional to 
$\mbox{cosech}(\pi q_{3} \lambda /2) $, which go to the constant  
$2 \lambda$ as $q_{3} \rightarrow 0$, 
while those proportional 
to $\mbox{sech}(\pi q_{3} \lambda /2) \rightarrow 0$ as 
$q_{3} \rightarrow 0$. This already indicates that dissipative terms 
involving ${\cal M}_{1}$, 
${\cal M}_{2}$, ${\cal M}_{3}$ and ${\cal M}_{4}$ will be more important than 
those coming from ${\cal M}_{6}$ and ${\cal M}_{5}$. 

It should be noticed that there is a natural cutoff 
momentum $k_{c}=2/(\pi \lambda)$ coming directly from the
wall structure. This can be converted to a cutoff frequency
$\omega_{c} = 2 c_{T}/ (\pi \lambda)$. For a wall of width $\lambda = 500 \AA$, 
and with a transverse sound velocity $c_{T} \sim 10^{4} \, \mbox{m/s}$, 
this gives a
cutoff frequency $\omega_{c} =  10^{11} \, \mbox{Hz} $. Walls with a 
very small
width will have an even larger $\omega_{c}$. This cutoff frequency is 
usually much larger than the bounce frequency $\Omega_{0}$, defined in Eq.
(\ref{om0}), at which the
bath must be cutoff from the truncation procedure. This makes it possible
to use $ \mbox{cosech}(\omega/\omega_{c}) \sim \omega_{c}/\omega$ and 
$ \mbox{sech}(\omega/\omega_{c}) \sim 1$. 

The exact form of the magnetisation profile is actually 
not too critical; we are 
mostly interested in the behaviour of these form factors at low-$q$. 
They are essentially an 
integral of the components of the non-uniform 
magnetisation out of the easy axis. The 
spatial variation of these terms is actually what defines the width of the
wall, so it is quite clear that the relevant form factors for the dissipative
dynamics of the wall behave as
\begin{equation}
\lim_{q \rightarrow 0} {\cal M}(q) \rightarrow {\cal B}_{\lambda} \lambda
\label{irre}
\end{equation}
where ${\cal B}_{\lambda}$ 
is a coefficient of $O(1)$ depending on the precise profile of
the magnetisation and $\lambda$ is the length scale over which the the 
magnetisation changes from one stable configuration to another. This should
be accurate provided that the curvature of the wall is weak, of course.

The two-point function of the phonon field is
decomposed as usual 
into a transverse and longitudinal part \cite{stoneham1}
\begin{equation} \langle u_{i}({\bf p},i\omega_{n})
u_{j}(-{\bf p},-i\omega_{n}) \rangle =
(\delta_{ij} -\hat{p}_{i}\hat{p}_{j})G^{T}({\bf p},i\omega_{n}) +
\hat{p}_{i}\hat{p}_{j}G^{L}({\bf p},i\omega_{n})
\end{equation}
with
\begin{equation} 
G^{T,L}({\bf p},i\omega_{n}) = \frac{1}{\rho_{v}}
\frac{1}{c_{T,L}^{2}p^2 + \omega_{n}^{2}}
\label{twopoint}
\end{equation}
where $c_{T}^{2} = \mu_{e}/\rho_{v}$ and 
$c_{L}^{2} = (\lambda_{e}+2\mu_{e})/\rho_{v}$ 
are
the transverse and longitudinal sound velocities. 

The effective action is found from a simple cumulant expansion. 
The first term, $\langle S_{I} 
\rangle$ is ignored: the 1-phonon term $\langle S^{1ph}_{I} \rangle$ 
is zero due to $\langle {\bf u} \rangle = 0$, and the 2-phonon term is 
independent of $Q(\tau)$, simply adding a contribution to the potential 
energy of the phonons (morphic effect). The ratio of the second-order 
magnetoelastic constants to the elastic energy is generally less than
$1 \%$ and we can simply ignore this renormalisation in the sound 
velocity. 
The first dissipative contribution to the 
effective action is thus $ \langle S_{I}^{2} \rangle /2 = 
\langle (S^{1ph}_{I})^{2} \rangle /2 + \langle (S^{2ph}_{I})^{2} \rangle /2$
which gives
\begin{eqnarray}
\Delta S_{eff} &=& \frac{1}{2} A_{ab} A_{cd} T \sum_{n} 
\sum_{{\bf q}} \int_{0}^{1/T} d\tau d\tau' 
e^{i \omega_{n} (\tau - \tau')}
e^{i q_{3} \cdot (Q(\tau)-Q(\tau'))}{\cal M}_{b}(-{\bf q}){\cal M}_{d}({\bf q}) 
{\cal G}_{ac} ({\bf q}, i\omega_{n}) \nonumber \\
& & + \frac{1}{2} R_{abc} R_{def} T^{2} \sum_{{\bf k}{\bf k}'} \sum_{n n'}
\int_{0}^{1/T} d\tau d\tau' e^{i (\omega_{n}+\omega_{n'})(\tau - \tau')}
e^{i (k_{3}+k_{3}') \cdot (Q(\tau)-Q(\tau'))} {\cal M}_{c}(-{\bf k}-{\bf k}') 
\nonumber \\ 
& & \times {\cal M}_{f}({\bf k}+{\bf k}') 
[{\cal G}_{ad} ({\bf k},i\omega_{n}) {\cal G}_{be} ({\bf k'}, i\omega_{n'})
+ {\cal G}_{ae} ({\bf k}, i\omega_{n}) {\cal G}_{bd} ({\bf k'}, i\omega_{n'})]
\label{entout}
\end{eqnarray}
where 
${\cal G}_{ab} \equiv \frac{1}{4} ( k_{a_{1}} k_{b_{1}} G_{a_{2} b_{2}}
({\bf k},i \omega_{n}) + \mbox{Perm} )$. The index $a \equiv (a_{1},a_{2})$ 
is the
pair of the indices introduced in Eq. (\ref{indices}) and 
Perm indicates the permutations $a_{1} \leftrightarrow a_{2}$ and
$b_{1} \leftrightarrow b_{2}$. A third term, of the form 
${\cal G}_{ab} {\cal G}_{de}$, is generated in the effective action, but it
does not couple to the position of the wall, simply corresponding 
to a further 
renormalisation of the sound velocity. Since it is not a dissipative term 
we do not include it in the effective action. It is now possible
to examine the different contributions to this action.

\subsection{1-Phonon Contributions to Action}

The 1-phonon part of the action is very simple.
Summing over the frequency, and separating the action into a transverse
and a longitudinal part, we obtain
\begin{equation}
\Delta S^{1ph}_{eff} = \Delta S^{1ph \, T}_{eff}+\Delta S^{1ph \, L}_{eff}
\end{equation}
\begin{eqnarray}
\Delta S_{eff}^{1ph, T} &=& \frac{1}{c_{T}} \frac{S_{w}}{\rho_{v}}  
\int_{0}^{\infty}
\frac{2 dq_{3}}{2\pi}
\int_{0}^{1/T} d\tau \, d\tau' \cos(q_{3}(Q(\tau)-Q(\tau'))) \, q_{3} \,
D(c_{T}q_{3},|\tau - \tau'|) \times \nonumber \\
& & \left[ A_{55}^{2} {\cal M}_{5}(-q_{3}) {\cal M}_{5}(q_{3}) + 
 A_{44}^{2} {\cal M}_{4}(-q_{3}) {\cal M}_{4}(q_{3}) \right]
\end{eqnarray}
\vspace{5mm}
\begin{eqnarray}
\Delta S_{eff}^{1ph, L} &=& \frac{1}{4c_{T}} \frac{S_{w}}{\rho_{v}} 
\int_{0}^{\infty}
\frac{2 dq_{3}}{2\pi} 
\int_{0}^{1/T} d\tau \, d\tau' \cos (q_{3}(Q(\tau)-Q(\tau'))) 
\, q_{3} \,   
D(c_{L}q_{3},|\tau - \tau'|) \nonumber \\
& & A_{3a}A_{3b} {\cal M}_{a}(-q_{3}) {\cal M}_{b}(q_{3})
\end{eqnarray}
where $D(\omega,\tau)$ is the boson propagator, defined in 
Eq. (\ref{boseprop}). 
This simple form of the effective action is obtained by taking 
advantage of the
fact that for a Bloch wall $q_{1}=q_{2}=0$. Next, inserting the
expressions for the ${\cal M}_{a}$'s, we obtain terms that contain only
the position of the wall, 
of
the form $\cos \omega (Q(\tau)-Q(\tau'))$. However, there will also 
appear terms 
like $\dot{Q}^2(\tau) \cos \omega (Q(\tau)-Q(\tau'))$ and 
$\dot{Q}(\tau) \dot{Q}(\tau') \cos \omega (Q(\tau)-Q(\tau'))$ 
coming from the velocity-dependent terms in Eq. (\ref{mdebut}) to 
(\ref{mfin}). 
We can rearrange $\dot{Q}(\tau)\dot{Q}(\tau')$ as 
$ (\dot{Q}^2(\tau) 
+ \dot{Q}^2(\tau)
 - (\dot{Q}(\tau) - \dot{Q}(\tau'))^{2})/2$ which brings  
a constant shift in the domain mass and a dissipative term, which depends on 
the velocity of the wall. We then obtain an effective action that can
be compared immediately with Eq. (\ref{dseff}), coming from the 
Caldeira-Leggett model \cite{leggettA} : 
\begin{eqnarray}
& & \Delta S_{eff}[Q,\dot{Q}] = \frac{1}{2} \int_{0}^{1/T} d\tau 
\int_{0}^{1/T} d\tau' (Q(\tau)-Q(\tau'))^{2} \alpha_{L} (\tau-\tau') 
\nonumber \\ &+& \frac{1}{2} \int_{0}^{1/T} d\tau 
\int_{0}^{1/T} d\tau' (\dot{Q}(\tau) - \dot{Q}(\tau'))^{2} \alpha_{T} 
(\tau-\tau') - \frac{\tilde{M}}{2} \int_{0}^{1/T} \dot{Q}^2(\tau)
\end{eqnarray}
where
\begin{equation}
\alpha_{L} = \frac{\pi}{4} \frac{S}{\rho} 
\frac{\lambda^{4}}{c_{L}^{7}} (A_{32}-A_{31})^{2} \int_{0}^{\infty}
d\omega \omega^{5} \mbox{cosech}^{2} \left( \frac{\pi \omega \lambda}{2 c_{L}}
\right) D(\omega,|\tau - \tau'|)
\label{long3d}
\end{equation}
\begin{equation} 
\alpha_{T} = \frac{\pi}{16} \frac{S}{\rho} 
\frac{\lambda^{4}}{c_{0}^{2} c_{T}^{5}} A_{44}^{2} \int_{0}^{\infty}
d\omega \omega^{3} \mbox{cosech}^{2} \left( \frac{\pi \omega \lambda}{2 c_{T}}
\right) D(\omega,|\tau - \tau'|)
\label{trans}
\end{equation}
\begin{equation}
\tilde{M} = \frac{1}{3} \frac{S}{\rho_{v}} \frac{\lambda}{c_{0}^{2}} 
\left[ \frac{1}{c_{L}^{2}} ( (A_{32}-A_{31})A_{33} -(A_{32}-A_{31})^{2}/2 )
+ \frac{8}{c_{T}^{2}} A_{44}^{2} 
\right]
\label{meff}
\end{equation}
This effective action is not completely equivalent to the 
Caldeira-Leggett action, due to the terms in $\dot{Q}$. This is of 
course a perfectly legitimate action, but its analysis is not simple. 
It however allows an easy identification of the dominant contribution to the 
dissipation. 
The normal dissipative 
term is of the superOhmic form ($J(\omega) \sim \omega^{3}$) as can be 
seen from Eq. (\ref{long3d}). This is what 
would be expected from one-dimensional one-phonon processes \cite{weisslivre}. 
It must also be expected in the case of dissipation for 
a Bloch wall, since the
wall restricts the phonon momentum in the plane of the wall to be zero
(see Eq. (\ref{64}).
The dissipative term coming from the
coupling of the phonons 
to the velocity is described by a spectral function linear in 
$\omega$, and 
looks at first sight to be of the Ohmic form.
However, since it depends on the velocity of the wall, its effect
is similar to superOhmic dissipation. 

In the transverse kernel $\alpha_{T}(\tau-\tau')$, 
there appears the Walker velocity $c_{0}$, defined in Eq. (\ref{vwalker}), 
and the transverse sound velocity
whereas only the longitudinal sound velocity appears in 
$\alpha_{L}(\tau-\tau')$.
In general, $c_{0} < c_{T} < c_{L}$ so that
depending on the value of the coefficient $A_{44}$, the term coming from the
velocity of the wall might be the most important. 
Secondly, for a crystal of cubic symmetry, $A_{32}=A_{31}$ and 
dissipation comes only from the coupling between the phonons and the
velocity of the wall. 
This can be traced back to our neglect of the influence of the coupling
to the lattice in the determination of the shape of the wall. Taking it
into account would give dissipation with a coefficient that would be smaller 
than the coefficient of the velocity-coupled 
dissipation by at least a factor of $A_{ab}/C_{cd}$, 
where $C_{cd}$ are the elastic constants. For nickel, this is a
reduction of $\sim 10^{-2}$, while for YIG it is $\sim 10^{-4}$,   
and it can be safely forgotten in both cases. 
This shows again
that the coupling to the velocity of the wall is quite important.
Let us then go back and perform the analysis by considering only the coupling
between the phonons and the velocity of the wall. 

\subsection{1-Phonon Coupling to the Wall Velocity}

We now look exclusively at a wall with a cubic crystalline structure or with
$c_{0} < c_{T}$ such that the dominant mechanism for 
dissipation is from the coupling of the phonons to the velocity of the wall. 
As the wall
is a one dimensional structure (in a flat wall approximation), 
one has the restriction 
$k_{1}=k_{2}=0$. The only relevant terms in the magnetoelastic Lagrangian are
thus
\begin{eqnarray}
L_{int} &=& \sum_{k_{3}} e^{i k_{3} Q(\tau )} [
A_{33} U_{3}(k_{3},\tau) {\cal M}_{3}(-k_{3}) +
A_{31} U_{3}(k_{3},\tau) ({\cal M}_{1}(-k_{3})+{\cal M}_{2}(-k_{3}))
\nonumber \\
&+&  A_{44} U_{4}(k_{3},\tau) {\cal M}_{4}(-k_{3}) 
+  A_{55} U_{5}(k_{3},\tau) {\cal M}_{5}(-k_{3}) ]
\label{cubique}
\end{eqnarray} 
${\cal M}_{3}$ and ${\cal M}_{1}+{\cal M}_{2}$ are proportional to 
$\dot{Q}^{2}$ while ${\cal M}_{4}$ and ${\cal M}_{5}$ are proportional to 
$\dot{Q}$. 
We are considering exclusively tunneling of the wall so that the 
excursions in position
can be restricted to second order in $Q$. This means that we can simply set
$e^{i k \cdot Q} = 1$ in Eq. (\ref{cubique}).
Furthermore, 
the first three terms in Eq.(\ref{cubique}) correspond to a renormalisation
of the wall's mass; this is very small and we neglect it here. The two 
remaining terms contain a coupling of the phonons to $\dot{Q}$. However, 
${\cal M}_{5} \rightarrow 0$ as $k_{3} \rightarrow 0$ while ${\cal M}_{4} 
\rightarrow \pi \lambda$ in the same limit. Thus, the  term in 
${\cal M}_{5}$ gives rise to higher powers of $\omega = c_{T} k_3$ 
than ${\cal M}_{4}$ 
and it is sufficient to consider only the coupling to ${\cal M}_4$. 
The effective
coupling between the phonons and the wall is mediated by the Fourier 
transform of the strain tensor $U_{4} = i k_{3} u_{2}(k_{3},\tau)$ (since
$k_{2}=0$) and the effective interaction Lagrangian is
\begin{equation}
L_{int} = \dot{Q}\sum_{k_{3}} \tilde{C}_{k_{3}}u_{2}(k_{3},\tau) 
\end{equation}
with the coupling constant
\begin{equation}
\tilde{C}_{k_{3}} = 4 i \pi \frac{A_{44}}{2 c_{0}} (k_{3} \lambda)^{3} 
\mbox{cosech} ( \pi k_{3} \lambda /2)
\end{equation}
Next, since the coupling is only to phonons in the $x_{2}$ direction,
and with the restriction $k_{1}=k_{2}=0$,  
we can use, as the new Lagrangian for the environment
\begin{equation}
L_{env} = \frac{1}{2} \sum_{k_{3}} \rho_{v} 
\dot{u}_{2}(k_{3},\tau) \dot{u}_{2}(-k_{3},\tau) 
+ C_{44} k_{3}^{2} u_{2}(k_{3},\tau) u_{2}(-k_{3},\tau)
\end{equation}
The mapping to the Caldeira-Leggett oscillator bath is clearly
$m_{k_{3}} = \rho_{v}/L_{3}$ and $\omega_{k_{3}} = c_{T} k_{3} \equiv 
(C_{44}/\rho_{v})^{1/2} |k_{3}|$.
The way to treat this problem is now straightforward. 
A total time derivative first needs to be introduced, in order to change the 
coupling between the wall and the phonons from $\dot{Q} u_{2}$ to 
$Q \dot{u}_{2}$.  
This coupling to the velocity of the 
environment can then be eliminated by a canonical 
transformation on the environment \cite{leggettA,leggettprb}. 
The resulting problem corresponds to a system
coupled by its position to a new set of oscillators, with the same $m_{k}$ and
$\omega_{k}$, but a new coupling constant $C_{k_{3}} = \omega_{k_{3}} 
\tilde{C}_{k_{3}}$ \cite{leggettprb}. 
The dissipative action is now in the usual 
Caldeira-Leggett form, with a spectral function
\begin{equation}
J^{(1ph)}_{\dot{Q}}(\omega) = \frac{\pi}{16} \frac{S_{w}}{\rho_{v}}
\frac{\lambda^{4}}{c_{0}^{2} c_{T}^{5}} A_{44}^{2} 
\omega^{5} \mbox{cosech}^{2} \left( \frac{\pi \omega \lambda}{2 c_{T}}
\right)
\label{jomegq}
\end{equation}

Thus the superOhmic $J(\omega) \sim \omega^{3}$ is recovered. This is the 
spectral function we shall use in the discussion of the dissipative 
effects of 1-phonon processes. We now turn to 2-phonon processes. 

\subsection{2-Phonon Contributions to Action}

Just as in the case of multiple magnons coupled to a domain wall
\cite{stamprev,stampprl}, the
extra phase space available for the interaction of pairs of phonons with
a
wall, means that 2-phonon couplings give a very different effect 
from 1-phonon couplings on the
wall dynamics. 
To start with, the summation over the 
magnetoelastic indices in Eq. (\ref{entout}) 
is obviously more complicated. Let us first consider 
a static wall, so that ${\cal M}_{1}=-{\cal M}_{2}$, ${\cal M}_{3}=0$, and
let us use only the components $R_{111}=R_{222}=R_{333}$ of the second order
magnetoelastic tensor. The summation is
then
\begin{equation}
|{\cal M}_{2}|^{2} [ R_{ab1} R_{de1}-R_{ab1}R_{de2}-R_{ab2}R_{de1}+
R_{ab2}R_{de2} ]  
[ {\cal G}_{ad}({\bf k},i\omega_{n}) {\cal G}_{be}({\bf k}',i\omega_{n'})
+ {\cal G}_{ae}({\bf k},i\omega_{n}) {\cal G}_{bd}({\bf k}',i\omega_{n'})]
\label{summationindex}
\end{equation}
We use the fact that $c_{T}<c_{L}$, 
which allows us to keep only the term $G_{T}(p,i\omega_{n})$ in 
Eq. (\ref{twopoint}) to simplify this
expression to
\begin{equation}
|{\cal M}_{2}|^{2} R_{111}^{2} G_{T}({\bf k},i\omega_{n}) 
G_{T}({\bf k}',i\omega_{n'}) 
[ k_{3}^{2} k_{3}'^{2} (k_{1}^{2} k_{1}'^{2}+ k_{2}^{2} k_{2}'^{2} )
+ k_{3}^{2} k_{1}'^{2} k_{2}'^{2} (k_{1}^{2}+k_{2}^{2}) +
 k_{3}'^{2} k_{1}^{2} k_{2}^{2} (k_{1}'^{2}+k_{2}'^{2}) ]
\end{equation}
Secondly, one must be
careful in performing the frequency summations. The two summations
(over $n$ and $n'$) 
should be decomposed as summations over the sum and the difference of the
frequencies, that is
\begin{eqnarray}
& & T^{2} \sum_{n=-\infty}^{\infty} 
\sum_{m=-\infty}^{\infty}
\frac{1}{\omega_{k}^{2}+\omega_{n}^{2}} \;
\frac{1}{\omega_{k'}^{2}+\omega_{m}^{2}}
e^{i(\omega_{n}+\omega_{m})(\tau-\tau')} = \nonumber  \\
&&-\frac{T}{2 \omega_{k} \omega_{k'}} 
\sum_{r=-\infty}^{\infty} e^{i\omega_{r}(\tau-\tau')}
\left( [1+n_{B}(\omega_{k})+n_{B}(\omega_{k'})]
\frac{\omega_{k}+\omega_{k'}}{\omega_{r}^{2}+(\omega_{k}+\omega_{k'})^{2}}
+  [n_{B}(\omega_{k})-n_{B}(\omega_{k'})]
\frac{\omega_{k}-\omega_{k'}}{\omega_{r}^{2}+(\omega_{k}-\omega_{k'})^{2}}
 \right)
\label{separation}
\end{eqnarray}
where $n_{B}$ is the boson occupation number, 
\begin{equation}
n_{B}(\omega_{k}) = \frac{1}{e^{\omega_{k}/T}-1}
\end{equation}

By this separation, two different contributions can be identified.
If we compare Eq. (\ref{separation}) 
 with Eq. (\ref{alphakernel}), it is clear 
that the wall can be considered to be  
coupled to two different oscillator baths. 
The first term on the right side of Eq.(\ref{separation}) corresponds to 
the simultaneous emission of two phonons followed by their re-absorption at a
later time, while the second term corresponds to the scattering of two phonons
off the wall. This latter process is not allowed 
for 1-phonon processes, 
due to energy-momentum conservation, but is perfectly possible once we 
consider 2-phonon processes. Of course, the scattering term requires phonons
to be already present and is thus non-existent at $T=0$ (we note that
precisely the same discussion applies to 
magnons, for 2-magnon processes-involving a
single bulk magnon- and 3-magnon processes, involving 2 bulk magnons
\cite{stamprev,stampprl}; the only complication there was the necessity
to include a single wall magnon as well).
The emission/absorption term can be mapped 
to a Caldeira-Leggett environment provided that we identify 
$\omega \equiv \omega_{k}+\omega_{k'}$. The low energy behaviour, 
$\omega \rightarrow 0$ then comes from the limit $\omega_{k}$, $\omega_{k'}
\rightarrow 0$, but looking back to Eq. (\ref{summationindex}), 
we see that this will implies
superOhmic dissipation with a very high power in $\omega$, due to the density
of states of the phonons going to zero as $\omega_{k} \rightarrow 0$. 
This term can be completely ignored since it will 
always be much smaller than the 1-phonon contribution. 

The scattering term however is very important as it 
represents Ohmic dissipation. The identification with a Caldeira-Leggett bath
requires $\omega \equiv \omega_{k}-\omega_{k'}$, so that the low energy 
properties of the bath come from $\omega_{k} \rightarrow \omega_{k'}$. 
Now, nothing special happens to Eq. (\ref{summationindex}) in this limit, 
so  the
effective density of states for 2-phonon scattering processes is extremely 
reduced. The resulting dissipation is Ohmic, and although the 
numerical coefficient in front of it may be small, the fact that it is Ohmic 
means that it may be the dominant dissipative process, especially for 
coherent processes \cite{leggettRMP,proko2}. 

The temperature dependence of the friction coefficient is determined by the
phonon density of states, and is thus different depending on the 
dimensionality of the phonons. We examine 1-, 2- and 3-dimensional
phonons and concentrate on the scattering term, since the
emission/absorption term results in superOhmic dissipation. 

\subsubsection{3-Dimensional, 2-Phonon Processes} 

It is now straightforward to obtain the friction coefficient, although the 
calculations are 
somewhat complicated by the anisotropy of the problem. 
Keeping in mind the restrictions $k_{1}+k_{1}'=0$ and $k_{2}+k_{2}'=0$ 
coming from the form factor of the magnetisation, it is useful to perform 
the change of variables
\begin{eqnarray}
\phi &=& \tan^{-1} k_{1}/k_{2} \nonumber \\
\rho &=& c_{T}(k_{1}^{2}+k_{2}^{2})^{1/2} \nonumber \\
\omega &=& c_{T}(k-k')=
c_{T} \left[ (k_{3}^{2}+\rho^{2})^{1/2}-(k_{3}'^{2}+\rho^{2})^{1/2} \right]
\\
\epsilon &=& c_{T}(k+k')= 
\frac{c_{T}}{2} \left[ (k_{3}^{2}+\rho^{2})^{1/2}+(k_{3}'^{2}
+\rho^{2})^{1/2} \right] \nonumber 
\end{eqnarray}
After doing the angular 
integrals, the most important contribution to the effective action (the
Ohmic one), can be isolated to give a dissipative action of the form
\begin{equation}
\Delta S_{eff}^{2ph} = \frac{T}{2\pi} \sum_{n} \int_{0}^{1/T} d\tau  
 \int_{0}^{1/T} 
d\tau' \int_{0}^{\omega_{D}} d\omega \, 
 \tilde{\eta}_{2ph}(Q(\tau)-Q(\tau'),\omega) \, e^{i \omega_{n} (\tau-\tau')}
\frac{\omega}{\omega_{n}^{2}+\omega^{2}} \sinh (\omega /2T)
\end{equation}
where the cutoff is taken as the Debye frequency, defined by 
$\omega_{D} = c_{T} k_{D}$ with $k_{D} \sim 1/a_{0}$, the inverse of the 
lattice spacing. The 
detailed information of the dissipation is contained in the function
$\tilde{\eta}_{2ph}(Q(\tau)-Q(\tau'),\omega)$. We are interested in the 
limit $\omega \rightarrow 0$ of the friction coefficient. 
After a further change of variables
$x = \epsilon/T$ and $ y  = T^{-1}(\epsilon^{2} - \rho^{2})^{1/2}$ 
we obtain
\begin{eqnarray}
& & \tilde{\eta}(Q(\tau)-Q(\tau'),0) = 
\left( \frac{R_{111}}{\rho_{v} c_{T}^{2}} \right)^{2}
\left( \frac{T}{c_{T}} \right)^{4} 
\int_{0}^{\omega_{D}/T} \frac{dx}{x^{4}} \frac{1}{\sinh^{2}(x/2)}
\nonumber \\
& & \int_{0}^{x} dy \, y \, 
( y^{6} - \frac{3}{2} x^{2} y^{4} + \frac{1}{2} x^{6} ) \, 
{\cal M}_{2}^{2}(2yT/c_{T}) \, 
\cos [ 2 y T(Q(\tau)-Q(\tau'))/c_{T}]
\label{etax}
\end{eqnarray}
Provided that 
$ k_{D} (Q(\tau)-Q(\tau')) \ll 1$ it is possible to expand
the cosine and keep only the quadratic term in $Q$.  
We are then in presence of conventional Ohmic Caldeira-Leggett dissipation. 
At temperature such that $T \ll \theta_{D}$, the Debye temperature
defined as $k_B \Theta_D = \hbar \omega_D$, the dissipation is then 
characterised by the friction coefficient
\begin{equation} 
\eta_{2ph}^{3d} = \frac{3}{5} \Gamma (7) \pi \hbar 
\left( \frac{R_{111}}{\rho_{v} c_{T}^{2}} \right)^{2}
S_{w} \lambda^{2} \left( \frac{k_{B} T}{\hbar c_{T}} \right)^{6} 
\label{eta3d}
\end{equation}
where we have reinstated $\hbar$ and $k_{B}$ to get the correct units,
$S_w$ is the area of the wall 
and $\Gamma (x)$ is the Gamma-function. 
Alternatively, it can be written in term of the 
Debye temperature $\Theta_{D}$ as
\begin{equation}
\eta_{2ph}^{3d} = \frac{108}{5} \Gamma (7) \pi^{5} 
\hbar \left( \frac{R_{111}}{\rho_{v} c_{T}^{2}} \right)^{2}
\frac{S_{w} \lambda^{2}}{a_{0}^{6}}
\left( \frac{T}{\Theta_{D}} \right)^{6}
\end{equation}
where $a_{0}$ is the lattice spacing of the material. 
The numerical factor comes from 
taking $k_{D}^{3} = 6\pi^{2}/a_{0}^{3}$. It shouldn't of course be taken too 
literally but nevertheless gives the right order of magnitude of the 
coefficient. 
Due to the presence of the temperature to the $6^{th}$ power, we do not 
expect the $2$-phonon processes to be relevant at really low temperatures, 
the question being of course: how low is sufficiently low!
We will consider this question below.

Notice that we did not include any terms coupling the phonons to the
velocity of the wall, as we did for 1-phonon processes. 
The reasons are twofold.
First, it is clear that the fact that we obtain Ohmic dissipation coming
from 2-phonon processes is not related in any way to the {\em form} of the
coupling between the phonons and the magnetisation. Therefore, including the
terms in $\dot{Q}$ would only result in temperature dependent 
superOhmic dissipation, a process much weaker than the 1-phonon terms
(remember that an apparent Ohmic dissipation in $\dot{Q}$ is
corresponds effectively to superOhmic dissipation). 
Secondly, the component of the second order magnetoelastic tensor required 
for such a coupling would be of the form $R_{555}$ and this is at least
four orders of magnitude smaller than $R_{111}$ (at least in bulk 
materials \cite{tremolet2,eastman}), and can be ignored.

\subsubsection{2-Dimensional Phonons and N\'eel Walls}

Since the N\'eel wall is essentially a 1-dimensional structure, the analysis
of 1-phonon dissipation in a thin film will be quite similar to the one in 
the bulk. The only difference is that there is now a large component of the
magnetisation in the direction of the wall motion. The form factors of the
N\'eel wall are thus
\begin{eqnarray}   
{\cal M}^{(N)}_{1} & = &  {\cal M}^{(B)}_{1} \; \; \; \;
{\cal M}^{(N)}_{4} = {\cal M}^{(B)}_{4} \nonumber \\
{\cal M}^{(N)}_{3} &\leftrightarrow& {\cal M}^{(B)}_{2}  \; \; \; \;
{\cal M}^{(N)}_{5} \leftrightarrow {\cal M}^{(B)}_{6}
\end{eqnarray}
Again the
contribution to the effective action contains longitudinal terms, and 
transverse  
terms coming from the velocity of the wall. Due to the structure of a N\'eel
wall, the longitudinal term is now
proportional to $(A_{33}-A_{31})^{2}$, which is non-zero even in the case
of a cubic structure. The transverse kernel, however, still contains the
Walker velocity $c_{0}$ and we can assume that it still gives the dominant
contribution to dissipation and can be used alone.
The analysis of the N\'eel wall with a velocity-coupling to the
phonons is
exactly analogous to what was done for the Bloch wall, and  the spectral
function is completely equivalent to Eq.(\ref{jomegq}):
\begin{equation}
J^{1ph}_{{\cal N}}(\omega) = \frac{\pi}{16} \frac{A_{w}}{\rho_{s}}
\frac{\lambda^{4}}{c_{0}^{2} c_{T}^{5}} A_{44}^{2}
\omega^{5} \mbox{cosech}^{2} \left( \frac{\pi \omega \lambda}{2 c_{T}}
\right)
\label{jomegn}
\end{equation}
where $A{w}$ is the cross-sectional 
area of the wall, $\rho_s$ is the surface
density and the constant $A_{44}$ is in units of $\mbox{J/m}^2$.

The procedure for analysing 2-phonon processes is similar to the 
3-dimensional case, 
but there are two differences. 
First, the density of states of the
phonons is reduced and this will reduce the the power of the temperature in
the friction coefficient. Secondly, the summation over the components of the
magnetoelastic tensor is
\begin{equation}
|{\cal M}_{3}|^{3} [ R_{ab1}R_{de1}-R_{ab1}R_{de2}-R_{ab2}R_{de1}+R_{ab2}
R_{de2} ] 
[ {\cal G}_{ad}({\bf k},i\omega_{n}) {\cal G}_{be}({\bf k}',i\omega_{n'})
+ {\cal G}_{ae}({\bf k},i\omega_{n}) {\cal G}_{bd}({\bf k}',i\omega_{n'})]
\end{equation}
If we assume $R_{111}=R_{222}$, then, due to the restriction
$k_{1}+k_{1}'=0$ coming from the $\delta-$function of the magnetisation
form factor, the whole summation is zero and there will be no dissipation
coming from 2-phonon processes. However, there is no particular 
ground for such an assumption,
as the second-order constants are almost unknown. In any case, if this is
true, then one will certainly couple to the other components of the
tensor, say $R_{112}$, $R_{122}$ and so on. We will assume that such a coupling
exists, so that we can write the friction coefficient for the motion of a
domain wall as 
\begin{equation}
\eta_{2ph}^{(2d)} \sim \Gamma (6) \hbar \left( \frac{\langle R \rangle}
{\rho_{s} c_{T}^{2}} \right)^{2} (A_{w} \lambda^{2})
\left( \frac{k_{B} T}{\hbar c_{T}} \right)^{5}  = 
\Gamma (6) (6 \pi^2)^{5/3} \hbar 
\left( \frac{\langle R \rangle}
{\rho_{s} c_{T}^{2}} \right)^{2} 
\left( \frac{A_{w} \lambda^{2}}{a_{0}^{5}} \right)
\left( \frac{T}{\Theta_{D}} \right)^{5}
\label{eta2d}
\end{equation}
where both $\langle R \rangle$ and 
$\rho_{s} c_{T}^{2}$ are in units of $\mbox{Jm}^{-2}$. 
 
\subsubsection{1-Dimensional Phonons in Magnetic Wires}

The 1-dimensional case is also straightforward to treat. Only longitudinal
phonons are present so the relevant components of the first- and 
second-order magnetoelastic tensors are $A_{3a}$ and $R_{33a}$
respectively. Without specifying the magnetisation
profile, we simply assume that there exists an average interaction coupling
$ \langle A \rangle \lambda$ and $\sim \langle R \rangle \lambda$ 
between the wall and the
phonons. The 1-phonon spectral function is then
\begin{equation}
J^{(1d)}_{1ph}(\omega) \sim \frac{1}{\rho_{l}}
\frac{\lambda^{4}}{c_{0}^{2} c_{L}^{5}} \langle A \rangle^2
\omega^{5} \mbox{cosech}^{2} \left( \frac{\pi \omega \lambda}{2 c_{L}}
\right)
\label{long1d}
\end{equation}
where we assumed that the dominant coupling was still to the velocity of
the wall. In cases where such a coupling is not present, the Walker
velocity $c_{0}$ should be replaced by the longitudinal sound velocity
$c_{L}$. 
The 2-phonon processes still give rise to Ohmic dissipation, 
with a friction coefficient 
\begin{equation}
\eta_{2ph}^{(1d)}  \sim \Gamma(5) \hbar \left( \frac{\langle R \rangle}
{\rho_{l} c_{L}^{2}} \right)^{2} \lambda^{2}
\left( \frac{k_{B} T}{\hbar c_{L}} \right)^{4} =  
\Gamma(5) (6 \pi^2)^{4/3}
\left( \frac{\langle R \rangle}
{\rho_{l} c_{L}^{2}} \right)^{2}
\left( \frac{ \lambda^{2}}{a_{0}^{4}} \right)
\left( \frac{T}{\Theta_{D}} \right)^{4}
\label{eta1d}
\end{equation}
with the expected reduction in the temperature dependence due to the reduction 
in the density of states. 

\subsection{Ohmic Coupling to Wall Chirality}
 
 The model of chirality tunneling, as proposed by Braun and Loss 
 \cite{braun1} and
 Takagi and Tatara \cite{tatara1} 
 was discussed in Section II. The models considered
 by
 these two groups are not completely equivalent. Braun and Loss
 consider
 a N\'eel wall in a wire while Takagi and Tatara use a Bloch wall. For
 definiteness, we concentrate on the Bloch wall model. The analysis can
 be trivially extended to a N\'eel wall.
  
  In the model, the magnetisation components are defined as
  $(m_{1},m_{2},m_{3})=(\cos \theta, \sin \theta \sin \phi, \sin \theta
  \cos \phi)$ with $m_{1}$ along the easy axis and $m_{3}$ normal to
  the
  plane of the wall. The angle $\theta = \theta (x_{3}-Q)$ describes a
  Bloch wall with centre at a position $Q$ along the $x_{3}$-axis.
  However,
  the wall is pinned and the dynamical variable is $\phi (\tau)$, with a
  tunneling event corresponding to a transition from
  $\phi = \pm \pi /2$ to $\phi = \mp \pi /2$.

We now consider the effect of phonons on this chirality tunneling.
   The magnetoelastic interaction keeps the form in Eq. (\ref{smgel1}),
   but two points must be noticed:
    
    (i) The position of the wall is time-independent
     
     (ii) The form factors of the magnetisation are now explicitly
     time-dependent, through $\phi(\tau)$. The relevant form factors are
     \begin{equation}
     {\cal M}^{B}_{2}({\bf q}) = - \frac{1}{2} (2\pi)^3 \delta (q_1 )
     \delta (q_2 ) q_{3} \lambda^{2}
     \mbox{cosech} ( \pi q_{3} \lambda /2) \sin^{2} \phi (\tau)
     \end{equation}
     \begin{equation}
     {\cal M}^{B}_{3}(q_{3},\tau )=  \frac{1}{2} (2\pi)^3 \delta (q_1 )
     \delta (q_2 )
     q_{3} \lambda^{2} \mbox{cosech} ( \pi q_{3} \lambda /2)
     \cos^{2} \phi (\tau)
     \end{equation}
     \begin{equation}
     {\cal M}^{B}_{4}(q_{3},\tau )=   \frac{1}{2} (2\pi)^3 \delta (q_1
     )
     \delta (q_2 )
     q_{3} \lambda^{2} \mbox{cosech} ( \pi q_{3} \lambda /2)
     \cos \phi (\tau) \sin \phi (\tau)
     \end{equation}
     We ignore ${\cal M}^{B}_{1}$ , since it does not depend on the angle $\phi
     (\tau)$,
     and although
     ${\cal M}^{B}_{5}$ and 
     ${\cal M}^{B}_{6}$ are non-zero, their effect is less important,
     since
     they go to zero as $q_{3} \rightarrow 0$, as discussed in Section
     IV-A 
     (cf. Eq. (\ref{irre})).
     There are of course no terms depending on the velocity of the
     wall, and
     the 1-dimensional character of the Bloch wall still restricts the
     momentum of the phonons in the plane of the wall to be zero.

It is now straightforward to integrate out the phonon field and to
obtain the dissipative action for the chirality
\begin{equation}
S_{eff} = \frac{\eta_{ab}^{\chi}}{4 \pi} \int_{0}^{1/T} d\tau
\int_{0}^{1/T} d\tau' T \sum_{n}
\int_{0}^{\omega_{D}} d\omega e^{i \omega_{n} (\tau - \tau')}
\frac{\omega^{2}}{\omega_{n}^{2}+\omega^{2}} \Phi_{a}(\tau)
\Phi_{b}(\tau)
\end{equation}
with $\phi_{0}(\tau)$ contained in the functions $\Phi_{a}(\tau)$, with
the values of $a$ and $b$ restricted to $a,b=2,3,4$, and
$\Phi_{2}= \sin^{2} \phi_{0}$, $\Phi_{3}= \cos^{2} \phi_{0}$
and $\Phi_{4}= \cos \phi_{0} \sin \phi_{0}$.
This action is not a simple functional of the tunneling variable, but
it
is nevertheless clear that the dissipation is Ohmic, with a
dimensionless coefficient $\alpha^{\chi} = \eta^{\chi}/\hbar$ given by
\begin{equation}
\alpha^{\chi}  \sim  \langle A \rangle^{2} \frac{\lambda^{2}}{\rho_{v}}
\frac{S_{w}}{ \hbar c_{L}^{3}}
\label{alpchi}
\end{equation}
If we consider a wall of surface $S_{w} = {\cal S} \AA^2$ in Ni
($\lambda \sim 500 \AA$, $A_{ab} \sim
10^8 J/m^3$ and $c_{L} \sim 10^3 m/s$), then we obtain
$\alpha^{\chi} \sim 10 {\cal S}$. Thus, even for a strictly
1-dimensional
wall, in a sample with all nuclear spins removed,
the coupling of 1-phonon processes to the chirality 
is obviously going to have a very severe effect on tunneling, and even
more on coherence. We discuss how severe this effect is in the next
section.

\subsection{Remarks on Power-Counting}

It is possible to explain the frequency dependence of the various
spectral functions obtained in this section by some simple power
counting arguments. 

Let us start with 1-phonon processes. 
The spectral
function describing the effect of isotropic acoustic phonons on the
tunneling of defects or interstitials in an insulator is of the form
$J(\omega) \sim \omega^{2+d}$ where $d$ is the dimensionality of the
system. This can be understood in the following
way. The square of the matrix element coupling the phonons to the defect
brings a power or $\omega$. The coupling is through a term of the form
$\exp (i {\bf k} \cdot {\bf q}(t) )$ where $k$ is the phonon's momentum
and
$q(t)$ is the position of the defect. Upon integration of the phonon
field, this term results in a factor $\cos (|{\bf k} \cdot {\bf q}(t)|)$
(cf Eq. (\ref{long3d}) and (\ref{trans}))  and it is clear that the
expansion of the cosine to
quadratic order in $(q(t)-q(t'))$ results in 2 more powers of
$\omega = c |{\bf k}|$ in the spectral function. Finally, the density of
states of the phonons brings a factor $\omega^{d-1}$.

With the magnetoelastic interaction, the matrix elements are the
magnetoelastic constants, but since the interaction is through the
strain tensor of the phonon field, this will also bring a power of
$\omega$ to the spectral function.
Thus, the spectral function of the phonons in a displacement tunneling
situation is clearly the results of the uni-dimensional character of the
Bloch wall, which reduces the effective density of states of the
phonons from $\omega^{d-1}$ to $\omega$, and this, independently of the
physical dimension of the substrate. 

Coming now to {\it chirality} tunneling, it is now obvious that 
the spectral function relating to the chirality tunneling should be 
Ohmic, since the phonons are effectively 1-dimensional, and there is no
coupling of the form $\exp (i k \cdot \phi(t) )$. Apart from numerical 
factors of $O(1)$, the different parameters entering the spectral 
function are understood easily from the spectral function of the
phonons 
in a displacement tunneling situation simply by removing two powers of 
$k = \omega/c_L$ from Eq. (\ref{long1d}) (and with the Walker velocity 
$c_0$ replaced by $c_L$).

The situation is more complex for 2-phonon processes. It was analysed in
detail by Kagan and Prokof'ev \cite{proko2} for the tunneling
and diffusion of defects in insulators. In $d$ dimensions, 
it was found that the Ohmic
dissipation resulting from 2-phonon processes was described by a
friction coefficient with temperature dependence 
$\eta (T) \sim T^{2+2d}$ (with ``transport effects'' included, as is
appropriate for isotropic phonons). Again however, this is the result
for a perfectly isotropic defect. A Bloch wall in a 3-dimensional
material imposes $k_1 + k_{1}'= k_2 + k_{2}' = 0$; the
problem is much more like the problem of magnons coupled to a wall
\cite{stamprev,stampprl}
(except that there one has the added problem of wall magnons). 
This then reduces
the phonon density of states by a power of 2, which then results in the
temperature dependence $\eta_{2ph}^{(3d)} \sim T^6$. In 2-dimensions,
the only restriction is $k_1 + k_{1}' =0$, and as such only one power of
the temperature will be missing with respect to the fully isotropic 
case, ie., $\eta_{2ph}^{(2d)} \sim T^5$. Finally, in 1 dimension, there
is no lowering of the density of states due to the shape of the wall
and we get $\eta_{2ph}^{(1d)} \sim T^4$. 

Now that the dissipative effects of the phonons are known, we go on to examine
the various dynamical processes of the wall.

\section{Tunneling of Domain Walls coupled to Phonons and Nuclear Spins}
\resetcounters

We now come to the main practical 
results of the present paper. In sections III
and IV we have seen that the main factors determining the tunneling
dynamics of a domain wall at low T (for $kT < \Delta$, the magnon gap)
will be (a) the coupling to any defect or other pinning potential (b)
the random hyperfine potential coming from nuclear spins, and (c)
dissipation coming from both phonons and nuclear spins. In the present
section, we will consider the case where the dissipative effects of
nuclear spins are small (ie., $\lambda_{I} \ll 1$, 
cf. Eq. (\ref{linumb})).
This will be the case for tunneling in wires of Ni or Fe (and
compounds of these). 

We are then left with the problem of a wall moving through a potential 
coming from both a pinning potential and the random hyperfine
potential (recall the latter is not small even if 
$\lambda_{I} \ll 1$), with dissipation coming from phonons only. Although
the random hyperfine potential varies slowly compared
to the ``bounce time'' for tunneling processes, it still changes very
quickly compared to any external field sweeps in an experiment.

At first glance the randomly-varying
$U_{hyp}$ appears to make a quantitative analysis of tunneling
impossible. However, the
situation is not so bad for Ni- and Fe-based materials, where the
hyperfine coupling $\omega_{0}$ is small, and the concentration $x$ of
nuclear spins is small. Even though 
$U_{hyp}(Q;\{ {\bf I}_{k} \} )$ may significantly
perturb  
$V(Q)$, it varies sufficiently slowly in time that during
a tunneling event, the total potential is essentially static. 
Since in any experimental run $\tilde{V}$ is not necessarily known, the main
effect of the $U_{Hyp}(Q)$ is then be to renormalise $\tilde{V}$
to a new value. A secondary effect is to slightly distort the
barrier shape from a quadratic/cubic shape in a way which can be discussed
quantitatively, for an ensemble of walls.
This must be done 
in terms of Eq.  (\ref{correlation}), and
corresponding higher moments. This kind of theory works best if the
change $\Delta U$ in the potential caused by the hyperfine potential
inside the wall satisfies
$|\Delta U|/\tilde{V} \ll 1$. This is well-satisfied for most Fe- and
Ni-based magnets.
For rare earth
magnets, 
$|\Delta U|/\tilde{V}$ may be very large indeed, and then an
analysis using the quadratic/cubic approximation will be almost
meaningless. 

Even though $|\Delta U|/\tilde{V} \ll 1$ for Ni- and Fe-based systems,
there are still rare large negative fluctuations of $\Delta U$ in
time. If the external field is swept slowly, the wall can tunnel long
before the ``bare'' pinning barrier is small. Thus 
the observed tunneling characteristics will depend on the sweep rate,
and, at slow sweep rates, the wall will cross over ``too early'' to
tunneling, from classical activation, and at too high a temperature.
 
A complete theory of the tunneling rate in this {\it dynamic} nuclear
potential turns out to be extremely difficult; the solution of the
dynamics in a {\it static} potential is difficult enough, even if one ignores
tunneling \cite{Bouchaud}. In the following we give an approximate
theory, valid for slow sweep rates, and for short $T_1$. Luckily these
conditions are appropriate to the Ni experiments. This is
described below; but first we discuss the tunneling in the complete
absence of nuclear spins. The tunneling in this case is a simple problem
of dissipative tunneling, entirely solvable with the results of section
IV; we also use these results to look at chirality tunneling and ``Bloch''
tunneling of walls.

\subsection{Wall Tunneling in Isotopically Purified Magnets}

We first consider the problem of wall displacement tunneling, ie., the
tunneling of the wall from some pinning centre, including the
dissipative 
influence of the phonons on this tunneling.
We then consider a system without any pinning centres (although possibly
with more microscopic defects or impurities), and analyse the
environmental influence on chirality tunneling and Bloch coherent
tunneling.

\subsubsection{Displacement Tunneling and Phonon Dissipation}

\begin{figure}
\epsfysize=5.5in
\epsfbox[0 100 100 600]{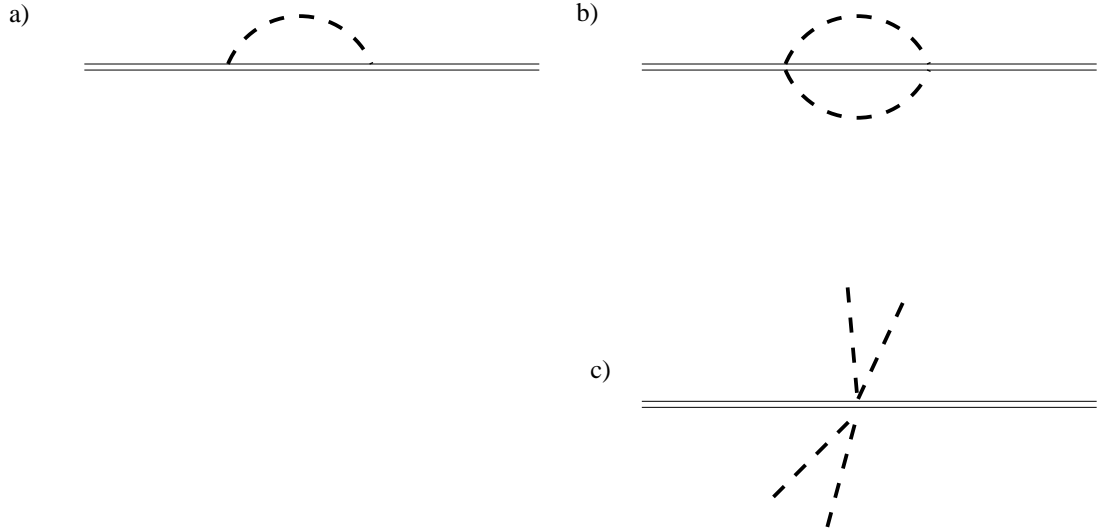}
\caption{Processes coming into the dissipative action of the domain
wall. 3-a) represent the 1-phonon processes, 3-b) represents the
simultaneous emission and reabsorbtion of 2 phonons and 3-c) is the
scattering process giving rise to Ohmic dissipation}
\label{phonons}
\end{figure}

When the defect potential is not perturbed by the nuclear spins, 
it is straightforward to obtain the corrections to the tunneling exponent
due to dissipation, since dissipation is fairly weak. We find that the 
total tunneling exponent $B(T)$ is given by  
\begin{equation}
B(T)=B_{0} \left[ 1 + 15 \alpha_{t}^{(d)}(T) + 
\frac{6}{\pi} \beta^{(d)}_{t} \left( 1- \left( \frac{T}{T_{0}} \right)^{4}
\right) \right] 
\label{bexponent}
\end{equation}
where $B_{0}$ is the tunneling exponent in the
absence of dissipation, given by Eq. (\ref{tunnelexp}). This expression
includes the effects of both Ohmic and superOhmic dissipation, and 
is valid for walls in bulk, films and wires, through the dimensionality
$d=1,2,3$ of the phonons. It is valid up to terms of order
$(T/T_0 )^6$, provided that one is not too close to the ``crossover
region'' between thermally activated and quantum barrier traversal
\cite{weisslivre,hanggi,grabert}. 
The dissipative processes entering in the
effective action are shown schematically on Fig. (\ref{phonons}). As
stated earlier, we only consider the superOhmic dissipation coming from
1-phonon processes and the Ohmic dissipation coming from the scattering
of a pair of phonons.
To obtain the result (\ref{bexponent}), we first 
 consider the bounce obtained in the absence of
 environment $Q(\tau) = Q_0  \, \mbox{sech} (\Omega_{0} \tau /2)$.
 Insertion
 of this solution into the effective action then yields the factor of
 $ 15 \alpha_{t}^{(d)}(T) + (6/\pi) \beta^{(d)}_{t}$. The temperature
 corrections to the exponent are found in the usual way by an expansion
 of the bounce around its zero temperature form
 \cite{weisslivre,hanggi,grabert}.
 Since the Ohmic dissipation due to 2-phonon processes is fairly weak
 and
 already strongly temperature dependent, it is not necessary to
 calculate
 the temperature correction brought by it. Ohmic dissipation should be
 taken into account in the non-zero temperature form of the bounce which
 is used to calculate the temperature corrections due to 1-phonon
 processes but again, since the dissipation is weak, these can be
 calculated, in a first approximation, as if only 1-phonon processes
 were
 present. The calculation is standard and gives the factor
 $ -(6/\pi^2) \beta^{(d)}_{t} (T/T_{0})^{4}$. As usual, the effect of
 dissipation is smaller at non-zero temperatures.

Thus all phonon effects depend on the size of the two dimensionless
parameters in (\ref{bexponent}).
Ohmic dissipation
comes from 2-phonon processes and the strength of this effect is parametrised
by the dimensionless coefficient 
$\alpha^{(d)}_{t}(T) = \eta_{2ph}^{(d)} (T)  
/2 M_{w} \Omega_{0}$, where $M_{w}$ in the mass of
the wall, Eq. (\ref{mdoring}), $\Omega_{0}$ is the oscillation 
frequency of the wall in the potential well, Eq. (\ref{om0}) and 
$\eta_{2ph}^{(d)}$, the
friction coefficient, is defined by the form of the spectral function
$J(\omega)=\eta \omega$ and given by Eq. (\ref{eta1d}),  
(\ref{eta2d}) and (\ref{eta3d}) for 1-, 2- and 3-dimensional phonons
respectively. Using the expressions for $M_{w}$ and $\Omega_{0}$, we
find
\begin{equation}
\alpha_{t}^{(d)}(T) = K_d \left( 
\frac{ \langle R \rangle^{(d)} }{ \langle C \rangle^{(d)} } \right)^{2}
\left( \frac{\lambda}{a_{0}} \right)^{3} 
\left( \frac{M_{0}}{H_{c}} \right)^{1/2}
\left( \frac{T}{\Theta_{D}} \right)^{3+d} \epsilon^{-1/4}
\label{alphat}
\end{equation}
with $\epsilon=1-H/H_{c}$ defined by the external magnetic field and the 
coercive field, Eq.(\ref{coer}), and $\langle R \rangle^{(d)}$ and 
$\langle C \rangle^{(d)}$ are the relevant second order magnetoelastic
constant and the elastic constant respectively. Thus 
$\langle R \rangle^{(d)}$ is in units of $\mbox{J/m}^{d}$, and 
$\langle C \rangle^{(d)} = \rho_d (c^{(d)})^2$, with 
$c^{(1)} = c_L$ and $c^{(d)} = c_T$ if $d=2,3$ and where 
$\rho_d$ is the density
appropriate to the dimensionality of the material. 
The constant $K_d$ includes the numerical factors of the spectral
functions Eq. (\ref{eta3d}), (\ref{eta2d}) and (\ref{eta1d}). For all
practical purposes, we can take the order of magnitude estimate
$K_d \sim 10^{2d+1}$. 
Notice that the power of the ratio between
$\lambda$ and $a_{0}$ is the same in all dimensions. This comes by using
a wall of surface $S_{w}=a_{0}^{2}$ in $d=1$ and $S_{w}=L_{1}a_{0}$,
with $L_1$ the cross-section of the wall in $d=2$. 

The usual 1-phonon processes give the expected superOhmic dissipation. 
We assume that the coupling between the phonons and the velocity
of the wall gives the dominant contribution to the dissipation, so that the
spectral function is given by Eq.(\ref{jomegq}). 
The strength of the dissipation is
given by the dimensionless parameter 
$\beta_{t} = \tilde{\beta} \Omega_{0}/M_{w}$ 
(coming from $J(\omega)=\tilde{\beta} \omega^{3}$) and can be expressed as
\begin{equation}
\beta_{t}^{(d)} = \frac{1}{2\pi} \left( 
\frac{ \langle A \rangle }{ \langle C \rangle^{(d)} } \right)
\left[ \frac{ \langle A \rangle \lambda
\gamma_{g}}{c^{(d)} M_{0}} \right] \left (\frac{H_{c}}{M_{0}} \right)^{1/2}
\epsilon^{1/4}
\label{betat}
\end{equation}
where $c^{(1)} = c_L$ and $c^{(d)} = c_T$ if $d=2,3$. Similarly, 
$\langle C \rangle^{(d)} = \rho_d c^{(d)}$, with $\rho_d$ the density
appropriate to the dimensionality of the material. 

We can now analyse the general feature of Eq. (\ref{bexponent}). The
salient feature is that there is a competition between 1-phonon and 
2-phonon processes. The 1-phonon part of dissipation is stronger at 
$T=0$ and decreases with increasing temperature while 2-phonon processes
cause an increase in dissipation with the temperature. Of course, 
everything depends on the relative strength of the dimensionless couplings
$\alpha_t$ and $\beta_t$. 
We now show that both $\beta_{t}$ and $\alpha_{t}$ are quite
small, with $\alpha_{t} < \beta_t $ in general.

Note first that $\alpha_t$ is strongly
reduced by the factor $(T/\Theta_D)^{3+d}$, and to a lesser extent by
the ratio of the second-order magnetoelastic constant to the elastic
constant. This reduction is however offset by the numerical factor 
$10^{2d+1}$, the ratio of the width of the wall to the lattice spacing
and by the ratio between the magnetisation and the coercive field.
Finally, there is a weak increase in dissipation as the height of the
tunneling barrier is reduced.  
The total value will however be extremely small. Let us consider some
parameters appropriate for Nickel (ie., $\lambda/a_{0} \sim 10^2$ and
$R/C \sim 10^{-1}$ and temperature such that $T/\Theta_D \sim 10^{-3}$.
The, we obtain $\alpha_{t}^{d} \sim 10^{-4-d} (M_{0}/H_{c})^{1/2} 
\epsilon^{-1/4}$. Thus, even for very small coercive fields and very
small tunneling barriers, the effect of 2-phonons processes can be
neglected, and this even in wires. However, in the region of $T \sim 1
K$, so that $T/\Theta_D \sim 10^{-2}$, the same estimate, in $d=1$ gives
$\alpha_{t}^{(1)} \sim 10^{-2} (M_{0}/H_{c})^{1/2}
\epsilon^{-1/4}$ and it is not obvious that such a term could be
neglected. Since the experiments of Giordano are performed above $1$ K,
the question of the relevance of the 2-phonon term remains open. 

The biggest contribution will thus come from the superOhmic dissipation
caused by 1-phonon processes. This dissipation is independent of
temperature, and decreases with decreasing coercive field and tunneling
barrier height. Again, using the parameters of Nickel (with the
first-order magnetoelastic constant $A \sim 10^8 J/m^3$), we obtain
$\beta_t \sim (H_c / M_0 )^{1/2} \epsilon^{1/4}$. The values of
$\beta_t$ should thus range between $0.01 - 0.1$ in an experiment.
This result also shows that the exact determination of $H_c$ becomes
quite relevant experimentally, as the value of $\beta_t$ depends mostly
on the ratio between $H_c$ and $M_0$ (at least in this simple minded
analysis for Nickel). 
In any case, even for a very strong
coercive field, it is unlikely that $\beta_t$ will become greater than 1
so that dissipation does not affect strongly the tunneling rate. The
dependence of the tunneling rate on the temperature is certainly the
most important effect arising from 1-phonon processes. In 
Fig. (\ref{tunnel}), we show the ratio $B(T)/B_0$ for a Bloch wall in a
typical Nickel sample, using 3-dimensional phonons, which allow us to
neglect 2-phonon processes. 

%\begin{figure}[h]
%\epsfxsize=6.5in
%\epsfbox[-110 150 593 653]{tunnel.ps}
%\caption{Temperature dependence of the tunneling exponent $B(T)$.
%We consider 3-dimensional phonons, with 2-phonon processes being
%irrelevant, and assume a value $6 \beta_t / \pi =0.2$.}
%\label{tunnel}
%\end{figure}

Finally, in YIG, the magnetoelastic constants are a lot smaller 
( $A/C \sim 10^{-6}$ and $R/C \sim 10^{-4}$ and we do not expect any
strong dissipative effect due to phonons, even in a temperature
dependence of the tunneling rate. 

\subsubsection{Bloch tunneling and Chirality Tunneling?}
 
 So far we have considered only the simple process of the tunneling
 depinning from a defect potential. However, suppose that there are no
 such pinning potentials. 
 What will then be the motion of the wall?
 One may consider one case, where, in
 absence of nuclear spins or dissipation, the wall would move quite
 freely (as in the case $\lambda \gg  a_{0}$). Alternatively, one may
 consider a situation where a very thin wall, with $\lambda/a_{0} \sim
 1-10$, tunnels through a periodic lattice potential (a variant on this
 uses an artificially imposed periodic lattice potential, acting on a
 light wall having $\lambda/a_{0} \gg 1$). 
  
  Let us first address the proposal that one might see coherent ``Bloch
  band'' motion of the walls in a periodic potential. Notice first just
  how restrictive the requirements are on such a process. One requires
  (a) almost perfect periodicity in the potential acting on the wall,
  such
  that deviations are much less than the bandwidth $\Delta_{0}$ for
  translational tunneling (b) extremely weak dissipative effects from
  phonons or other bosonic environments, and (c) no possibility of any
  phase decoherence - by far the most serious source of which 
 is environmental spins (topological decoherence).
   
   The requirement of near perfect periodicity in the total static
   potential wall potential $V_{W}(Q)$ seems
   almost
   impossible to satisfy, at least in the foreseeable future.
    The basic problem is that the bandwidth $\Delta_{0}$ will be
    extremely
    small, even for very small and light walls. Thus, Braun and Loss
    \cite{braun2,braun1} 
    propose
    the example of a YIG wire with cross-section $100 \AA \time 100
    \AA$,
    with a wall having $\lambda \sim 400 \AA$; for this very light wall
    they
    estimate $\Delta_{0} \sim 80 mK$ (and a crossover to tunneling
    $T_{0} \sim 50 mk$).
    Thus, to satisfy the criterion of near periodicity in the
    potential,
     one requires that extrinsic
     corrections to $V_{W}(Q)$  (coming from non-uniformities in the
     wire
     cross-section, dislocations, defects, etc), be much smaller than
     $80
     mK$. However for this example the wall surface energy $\sim 10^{-4}
     J/m^{2} \sim 7 \times 10^{18} K/m^{2} \equiv 0.07 K/\AA^{2}$, so this
     is a
     very tough requirement - the displacement of a single atom in the
     wire will cause a 
     local change in $V_{W}(Q)$  considerably {\bf larger} than
     $\Delta_{0}$!! In reality it is difficult at present to manufacture
     magnetic wires with better than a 5-10 per cent non-uniformity in
     wall cross-section, so that the fluctuations in the potential will
     be over energy scales some 3 orders of magnitude larger than this.

\begin{figure}
\epsfysize=5.5in
\epsfbox[0 100 100 600]{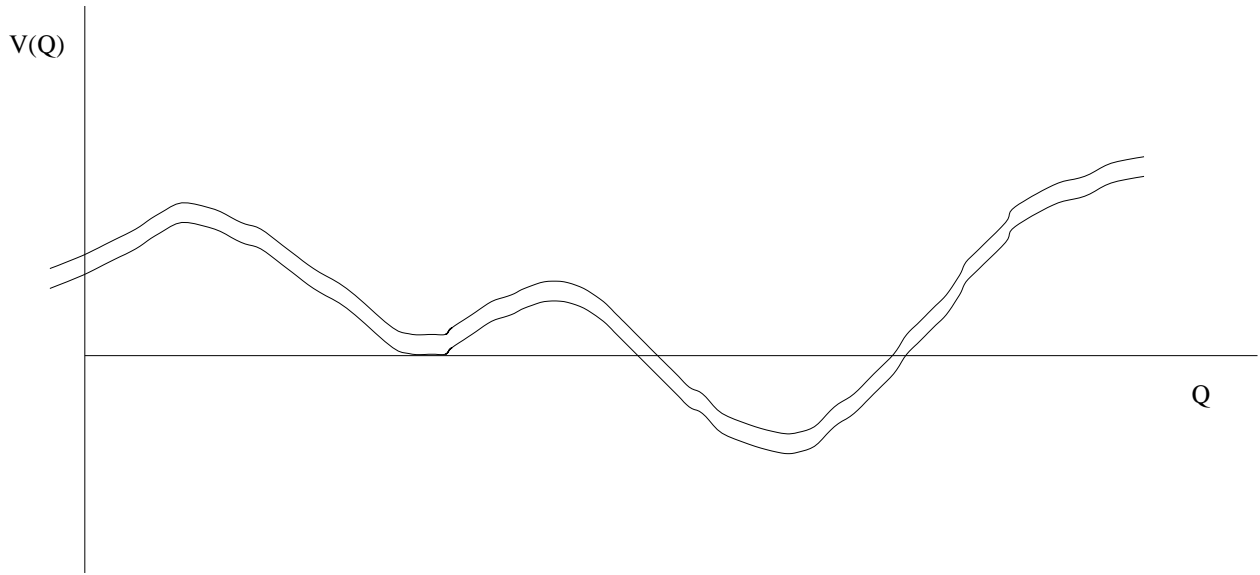}
\caption{Distortion of the putative Bloch band, caused 
by imperfections in the
background periodic potential. This picture is meaningful if the variation
caused by imperfections is over length scales long compared to the lattice
spacing.}
\label{band}
\end{figure}

Thus the potential felt by the wall will realistically be like that
shown in Fig. \ref{band}, with some very small periodic 
component superposed.
Problems like this were considered in great detail in studies of the
quantum diffusion of defects in solids (particularly muons).
\cite{KagProk} The actual motion of the wall will be one of quantum
diffusive motion between sites (eg., neighbouring potential wells), in
which a thermal bath takes up the energy difference
between sites (here via inelastic emission of phonon pairs, this being
the dominant Ohmic process at low $T$).

%\begin{figure}
%\epsfysize=5.5in
%\epsfbox[0 100 100 600]{band.ps}
%\caption{Distortion of the Bloch band due to imperfections in the
%crystal structure}
%\label{band}
%\end{figure}

Suppose
that at $t=0$ the wall is found in a potential well as shown on 
Fig. (\ref{band}).
Then, if $\delta Q$ is the distance
between this potential well and a nearby well with a lower minimum, the
wall will tunnel at a rate given by quantum diffusion theory as
\cite{KagProk}
\begin{equation}
W= \frac{2(\Delta(\delta (Q))^{2}
\Omega_{2ph}(T)}{\xi^{2}+\Omega_{2ph}^{2}(T)}
\frac{\xi/T}{1-e^{\xi/T}}
\end{equation}
with
\begin{equation}
\Omega_{2ph}(T) = T \eta_{2ph}(T) (\delta Q (\xi) )^2 /\hbar
\end{equation}
where $\eta_{2ph}(T)$ is the 2-phonon Ohmic coupling calculated in
section IV-C, and $\xi$ is the energy difference between the bottom of
the 2 wells; the tunneling matrix element
$\Delta(\delta (Q))$ is that between the 2 wells (notice that
$\Delta(\delta (Q))$ decreases exponentially as $\delta Q$ increases, so
that in most cases $\delta (Q)$ will be the distance between
neighbouring potential wells of the periodic potential). Without more
information about the nature of the background random potential, one
cannot say much more here except that the wall will, at low $T$, move
quantum diffusively towards the bottom of the nearest potential well of
this potential.

Turning now to the possibility of chirality tunneling, one notes again
the requirement that dissipation and decoherence be absent, as well as
a complete energetic symmetry between the 2 chiral states. The
requirement of energetic symmetry is still going to be extremely
difficult to satisfy, and we also now have the very strong Ohmic
coupling of single phonons to the chirality (Eq. (\ref{alpchi})). From
the example given after Eq. (\ref{alpchi}) we see that the
dimensionless Ohmic coupling $\alpha_{\chi} \gg 1$ in almost all cases.
This means, according to the standard results for Ohmic $T$-independent
coupling, \cite{leggettA,leggettRMP} that (a) coherence will be
suppressed completely (indeed the chirality variable will be {\it
localized} at low $T$ by the coupling to the phonons!), and (b) that the
incoherent tunneling rate between the 2 chiralities will be suppressed
by many orders of magnitude- so much so that we believe that for any
reasonably-sized wall it can be neglected (it is probably more to the
point to ask why {\it thermally activated} transitions between chirality
states has not been observed at higher temperatures).

This summarizes the results for an isotopically purified system. We see
that the physics of wall displacement tunneling in an isotopically
purified system is only weakly perturbed by phonons, in a way easily
evaluated theoretically; thus it would be very interesting to see
experiments performed in such ideal conditions. We expect the results to
be very different in the presence of nuclear spins, as we now see.

\subsection{Wall Tunneling in Dynamic Hyperfine field}

We first analyse the tunneling for a Bloch wall; the results for a
N\'eel wall are then obvious. The
quasi-static potential experienced by the domain wall will be a sum of
defect and random hyperfine fields:
\begin{equation}
V_{W}(Q) = V(Q) + U_{hyp}(Q;\{ {\bf I}_{k} \} )
\label{totalv}
\end{equation}
where $V(Q)$ is given by Eq. (\ref{pinpot}) and
$U_{hyp}(Q;\{ {\bf I}_{k} \} )$ by Eq. (\ref{pinpotnucspin}).
Over time scales $\Omega_{0}^{-1}$, we see 
$U_{hyp}(Q;\{ {\bf I}_{k} \} )$ is indeed
static. 

In addition, the wall is coupled dynamically to phonons, via
Eq. (\ref{mgel}), and
to the nuclear spins fluctuations, via (\ref{nusp}).
 
At first glance it might seem that in systems like Ni, where the
nuclear spins concentration $x \sim 0.01$, and $\omega_0 = 28.35 \, 
\mbox{MHz}$ is very small, we might begin by ignoring the hyperfine
field in favour of the ``bare potential'' $V(Q)$. Consider the ensemble
averaged value of 
$\Delta U_{12} = {\cal C}_{U_1 U_2}^{1/2}$ between points on either side
of the tunneling barrier, here $Q_1$ is the entry point for tunneling
and $Q_2$ is the exit point; recall (Eq. (\ref{cul0})) that 
$Q_1 - Q_2 = \lambda \epsilon^{1/2}$, in terms of $\epsilon = 
1 - H_e / H_c$. We may estimate that 
\begin{equation}
\frac{|\Delta U|}{\tilde{V}} = \frac{\omega_{0} I}{\mu_{0} \gamma_{g} H_{c}}
\left( \frac{x}{N_{0}} \right)^{1/2} \epsilon^{-5/4}
\label{fracU}
\end{equation}
where $\tilde{V}$ is the barrier height at an applied field $H_e$ (cf., Eq.
(\ref{barrier})). 
For a Nickel wire, using the numbers in section II (cf., discussion
following Eq. (\ref{215}), one finds that $|\Delta U|/\tilde{V}
\sim 0.01$ only.
Since in any experiment $\tilde{V}$ is not known, the
main
effect of this small $U_{Hyp}(Q)$ will be to simply renormalise 
$\tilde{V}$
to a slightly different value, which is not directly observable, and
which does not apparently change the physics very much (of course the
same is not true for rare earth magnets, where we can easily verify the
typically $| \Delta U| / \tilde{V} \gg 1$). 

However even for Ni wires the above argument ignores the fact that if
the applied field is static, or being swept only very slowly, 
then the actual tunneling of the wall can be controlled
by the relatively rare but strong downward (ie., negative) temporal
fluctuations in $U_{hyp}(Q;\{ {\bf I}_{k} \} )$ occurring in the
tunneling barrier region. 

A general treatment of this problem is beyond the scope of this paper.
Here we will give an approximate treatment. Let us first consider the
case of static external field (ie., zero sweeping rate), with some fixed
small value of $\epsilon$. In this case we shall define 3 characteristic
time scales for the problem, viz., 

(i) the tunneling traversal or bounce time $\Omega_{0}^{-1}$, which will
be very short (typically, $\Omega_0 > 10^9 \, \mbox{Hz}$). 

(ii) The characteristic time $^{\bot}T_{Diff}(\epsilon)$ for a
fluctuation in the hyperfine field $U_{hyp}$ to diffuse into or out of
the barrier; since the barrier width is $\sim \lambda \epsilon^{1/2}$,
we have
\begin{equation}
^{\bot}T_{Diff}(\epsilon) \sim T_2 \frac{\lambda^2}{4a_{0}^{2}} \epsilon
> T_2 
\label{cond0} 
\end{equation}

(iii) Finally, the experimental tunneling time $\tau$, ie., the inverse
of the tunneling rate $\Gamma$ 
at a given $\epsilon$, in the presence of the
fluctuating hyperfine potential; we shall estimate this below. 

Let us consider the regime where 

\begin{equation}
\Omega_0 (\epsilon) \gg T_2 
\label{cond1}
\end{equation}
\begin{equation}
^{\bot}T_{Diff}(\epsilon) \gg \tau^{-1} (\epsilon)
\label{cond2}
\end{equation}

The first inequality simply demands that the hyperfine potential be
static during tunneling traversal - since $T_{2}^{-1} \sim 10^4 - 10^6
 \, \mbox{Hz}$, this is always satisfied. The second inequality
corresponds to the assumption that the wall always has enough time to
sample enough tunneling paths (made available to it by fluctuations in
$U_{hyp}$ through the space of possible potentials), that it will sample
nearly ``optimal'' realisations of $U_{hyp}$ (ie. those giving the maximum
tunneling rate, or near to it). 
We shall see below when this
assumption is consistent with our final answer for $\tau^{-1}
(\epsilon)$. Once Eq. (\ref{cond1}) and 
(\ref{cond2}) are obeyed, we can estimate 
$\tau^{-1} (\epsilon)$  by functionally averaging over possible
realisations of $U_{hyp}$, with a tunneling action appropriate to each,
to get 
\begin{equation}
\tau^{-1} (\epsilon) \sim \Omega_0 (\epsilon) 
\int {\cal D} U(Q) {\cal P} ( \{ U(Q) \} ) e^{-S( \{ V_W (Q) \},
\dot{Q})}
\label{probrate}
\end{equation}
where ${\cal P} ( \{ U(Q) \} )$ is the probability for a particular
realisation $U(Q)$ of the hyperfine potential to occur, and 
$S( \{ V_W (Q) \}, \dot{Q})$ is the action for the tunneling
trajectory through the total potential 
$V_W (Q) = V(Q)+U(Q)$ (cf., Eq. (\ref{pinpotnucspin})). 
This equation makes no attempts
to get pre-exponential factors right. Notice that 
Eq. (\ref{probrate}) is independent of
$T_2$, because of condition (\ref{cond2}). Thus the essential
simplifying assumption in Eq. (\ref{probrate}) is simply that if 
$\tau (\epsilon)$ is sufficiently long compared to 
$^{\bot}T_{Diff}(\epsilon)$,
we do not have to worry about the ``waiting time'' required for $U(Q)$
to first approach an optimal realisation, everything boils down to the
measure of such optimal configurations.

The assumptions Eq. (\ref{cond0}) - (\ref{cond2}) are probably satisfied
in most wall tunneling problems. However, even with these assumptions,
Eq. (\ref{probrate}) is not easy to evaluate, it is a complicated path
integral over different ``paths'' $U(Q)$ between $Q_1$ and $Q_2$. We
intend elsewhere to give a thorough analysis of  (\ref{probrate}). Here
we shall estimate the change in $\tau (\epsilon)$ brought about by the
fluctuating $U_{hyp} (Q)$ in the weak-coupling regime, where 
$| \Delta U |/ \tilde{V} (\epsilon) \ll 1$, appropriate to Ni- and
Fe-based systems, and show that the change is not important enough to
explain the crossover temperature discrepancy reported by Giordano et
al.

To make this estimate, we introduce a ``typical'' potential 
$U_{\alpha} (Q)$, having the form
\begin{equation}
U_{\alpha} (Q) \sim \alpha E_0 (Q/Q_0)^{1/2}
\label{ansatz}
\end{equation}
for $0 \le Q \le Q_0$, where the bare pinning potential $V(Q)$ has a
minimum at $Q=Q_0$, and the tunneling end-point is at $Q=Q_0$. The value
$\alpha=1$ refers to the ensemble-averaged ``gaussian half-width''
value, ie., $E_0 = | \Delta U_{12} |$, so that 
\begin{equation}
E_0 = \omega_0 s I N^{1/2} (\epsilon)
\end{equation}
where $N(\epsilon) \sim N(\lambda) \epsilon^{1/2}$ is the number of
nuclear spins swept by the wall in going from $Q=0$ to $Q=Q_0$, and
$N(\lambda) = x (S_w \lambda /a_{0}^{3} )$ as before. The ansatz 
Eq. (\ref{ansatz}) allows us to reduce (\ref{probrate}) to a 
1-dimensional integral:
\begin{equation}
\tau^{-1} (\epsilon) \sim \Omega_0 (\epsilon) \int_{-\infty}^{\infty} 
\frac{d\alpha}{\sqrt{2\pi}} e^{- \alpha^2/2} \exp \left( 
 - \frac{1}{\hbar} \int dQ (2 M_w V_{w}^{\alpha} (Q))^{1/2} \right)
\label{simple}
\end{equation}
and where the total potential 
$V_{w}^{\alpha} (Q)$ is given by the sum of $V(Q)$
(evaluated in quadratic/cubic approximation) and $U_{hyp} (Q)$
(approximated by $U_{\alpha} (Q)$ in (\ref{ansatz})), ie., 
\begin{equation}
V_{w}^{\alpha} (x) = \frac{27}{4} \tilde{V} (\epsilon) (x^2-x^3) +
\alpha E_0 |x|^{1/2}
\end{equation}
where $x=Q/Q_0$.

Eq. (\ref{simple}) is a drastic simplification of (\ref{probrate}), but
it allows us to extract the exponent in the tunneling rate in the limit
where $E_0 \ll \tilde{V} (\epsilon)$, so that the last term in 
$V_{w}^{\alpha} (Q) $ is a small perturbation. If we write the bare
tunneling amplitude exponent $B_0 (\epsilon) \sim (M_w \tilde{V}
(\epsilon) )^{1/2}$ (this is equivalent to Eq. (\ref{tunnelexp}) or 
Eq. (\ref{215}), within a
constant $\sim O(1)$), then we can write the exponential in
(\ref{simple}) as $\exp [- B(\epsilon, \alpha)/\hbar]$, where 
\begin{equation}
B(\epsilon,\alpha) = B_0 (\epsilon) \left( 1 + 
 \left( \alpha \frac{E_0}{\tilde{V}(\epsilon)} \right)^{2/3}
\; sign \;\alpha +
\frac{3}{2} \alpha \frac{E_0}{\tilde{V}(\epsilon)} + 
O( (\alpha E_0 /\tilde{V})^x ) \right)
\label{nucspinexp}
\end{equation}
where $x > 1$. 
For $E_0 / \tilde{V}(\epsilon) \ll 1$, the second term in 
Eq. (\ref{nucspinexp}) dominates over the third one in the integration
in (\ref{simple}). Doing this integral by steepest descents yields a new
tunneling rate 
\begin{equation}
\tilde{\Gamma} \sim \Gamma_0 (\epsilon) e^{- \tilde{B} (\epsilon) }
\end{equation}
\begin{eqnarray} 
\tilde{B} (\epsilon) & = & B_0 (\epsilon) - \Delta B (\epsilon)
 \nonumber \\
& \sim & B_0 (\epsilon) - c (E_0/\tilde{V} (\epsilon) ) B_{0}^{3/2}
(\epsilon)
\end{eqnarray}
where $\Gamma_0 (\epsilon)$ is the bare tunneling rate 
(Eq. (\ref{tunnelexp})), and
$c =  \sqrt{3/2} -2/9 \sim 0.6$. Notice that whereas 
$B_{0} (\epsilon) \sim \epsilon^{5/4}$ (Eq. (\ref{215})), we have 
$\Delta B (\epsilon) \sim \epsilon^{3/8}$. The decrease in the relaxation
time $\tau (\epsilon) \equiv \tilde{\Gamma}^{-1} (\epsilon)$ is entirely
due to the occasional downward fluctuations in $U_{hyp} (Q)$. However,
we emphasise that in the regime of validity of our estimation of 
$\Delta B (\epsilon)$, ie., for $E_0/\tilde{V} (\epsilon) \ll 1$, we
expect $\Delta B (\epsilon) / B_0 (\epsilon) \ll 1$ as well (for values
of $\tilde{B} (\epsilon)$ corresponding to observable relaxation times).
In particular, for Ni-based magnetic wires, we do not expect a large
effect on $\tau (\epsilon)$. 

This concludes our analysis of the influence of nuclear spins on wall
tunneling. We conclude that in Ni- and Fe-based systems, their influence
is fairly small. However, in rare earth systems, where 
$E_0/\tilde{V} (\epsilon) \gg 1$, it is clear that their effects on
both tunneling and the classical wall dynamics will be very large, and
will require a more sophisticated theory, starting from the effective
interaction Hamiltonian derived in section III (Eq. (\ref{nucspinlong})
and (\ref{nucspintrans}).

Finally, we note that the nuclear spins will also adversely affect any
Bloch tunneling, since $U_{hyp} (Q)$ further disturbs any periodicity in
the potential. For Ni- and Fe-based systems, this effect is certainly
much smaller that that caused by disorder and non-uniformities in the
wire cross-section; but for rare earth magnetic wires it could easily be
larger.

\section{Discussion: Relation to Experiments }

Let us now put together what we have learned from the results in
sections III-V. With the assumption that for temperatures well below the
magnon gap energy, the only significant environmental effects on domain
wall tunneling can come from phonons and nuclear spins, we have set up a
fairly complete theory for these effects in the limit where the nuclear
spin effects are weak (the limit appropriate to Ni- and Fe-based
magnets). We find that in this limit the phonon effects can be evaluated
more or less exactly at these low energies, and involve 3 main effects -
a superOhmic 1-phonon coupling to the wall velocity, a strongly
$T$-dependent Ohmic coupling to the wall position (both of which are
usually quite weak), and a very strong Ohmic 1-phonon coupling to the
wall chirality. For a system isotopically purified of nuclear spins,
these calculations suffice to determine completely the low-$T$ tunneling
behaviour; the ``displacement tunneling'' of the wall is weakly modified
at finite $T$ from that originally calculated by Stamp
\cite{stamprev,stampprl}. On the other hand chirality tunneling is
suppressed, and both it and ``Bloch tunneling'' seem to be practically
unobservable, even for extremely small domain walls.
Addition of nuclear spins makes the problem
much harder, but for Ni- and Fe- based sytems the effect of the
dynamic fluctuating longitudinal hyperfine field 
on wall displacement tunneling is small enough that
the change in the tunneling exponent can be estimated. One finds an
increase in the tunneling rate over that expected without nuclear spins,
caused by ocasional strong negative fluctuations in the hyperfine
potential.

\begin{figure}
\epsfxsize=6.5in
\epsfbox[-110 150 593 653]{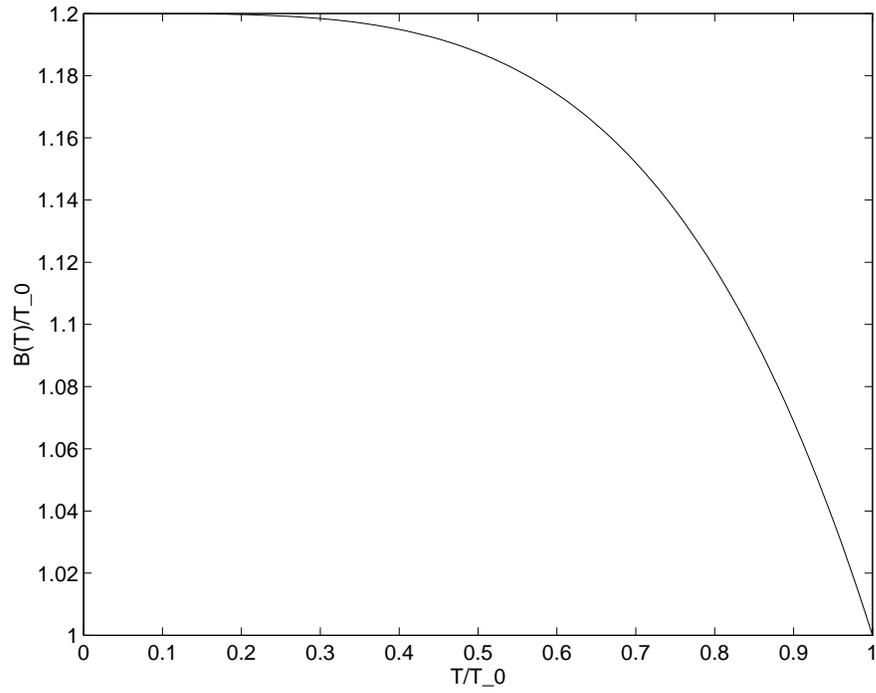}
\caption{Temperature dependence of the tunneling exponent $B(T)$,
in the absence of nuclear spins.
We show the result for 3-dimensional phonons, with 2-phonon processes being
irrelevant, and assume a value $6 \beta_t / \pi =0.2$.}
\label{tunnel}
\end{figure}

The experimental implications for isotopically purified systems have
already been discussed in section V.A. We summarize the key result in
Fig. (\ref{tunnel}), which shows a typical T-dependence of the rate 
exponent for
wall displacement tunneling from a pinning potential. It would be of
some interest to test this result experimentally on, eg., an
isotopically purified Ni magnetic wire, since the theory contains no
adjustable parameters once the small oscillation frequency $\Omega_0$,
the wall coercive field $H_c$, and
the magnetostrictive contants are known. 

For the case where the system contains naturally-occurring nuclear
spins, a large number of experimental results exist already. The work 
by Giordano et al. on nominally single domain walls in Ni
wires, includes (a) 
measurements of the tunneling rate $\Gamma$,
at low $T$, and the quantum/classical crossover temperature $T_0$, on
various wires \cite{giordtun} (b) measurements of the escape field
statistics over a range of temperatures \cite{giordesc}, and (c) microwave
measurements over a range of applied fields and temperatures, which 
give some evidence for level quantization of the trapped wall
states \cite{giordres}. In addition to these results there are many
experiments on more complex wall systems, which give various kinds of 
evidence for tunneling \cite{Uehara87,Paulsen91,Zhang92,multiwall}.
These experiments have been done at a variety of sweep rates, and for a
wide variety of materials. 

We may make a number of remarks concerning these
results, on the basis of the present theory, as follows:

(i) {\it Quantum/Classical Crossover}: It is noticeable
that some of these experiments show a crossover temperature  
$T_0$ considerably higher than the predicted \cite{stampprl,stamprev} value.
In the experiments $T_0$ has usually been determined by looking
for a low-$T$ crossover to a plateau, in either the relaxation rate or the
escape field; more sophisticated analyses look at the distribution of
escape fields. One typically expects $T_0$ to {\it decrease} with
increasing dissipative coupling to 
an oscillator bath, although how it does so,
and the actual $T$-dependence of the transition rate, obviously depend
on the nature of the bath (we are at a loss to explain why some
experimentalists insist on using formulas appropriate to {\it Ohmic}
baths to describe the crossover; this is never correct unless the low-$T$
dissipation is dominated by electrons, which probably only happens for
extremely thin walls, if at all). For
isotopically purified systems it is clear from Fig. (\ref{tunnel}), that the
simple estimate that $T_0 \sim (\Omega_0/ 2 \pi)$, is reasonably
accurate.
However adding the nuclear spins makes the crossover much more complex.
Suppose first the applied field is static (or being swept extremely
slowly)- then the estimates in the previous section apply. We see
immediately that for a given temperature, 
the distribution $P(H)$ of measured escape
fields will be now be affected by both quantum/thermal
fluctuations {\it and} by the fluctuations of $U_{hyp}(Q)$ in the barrier
region, ie., there will be a $T$-independent extra spread in 
$P(H)$. A rough estimate of this extra
spread is obtained from the curvature of $B(\epsilon, \alpha)$ about
its minimum in $\alpha$; this gives an extra spread $\sim E_0$ in bias
$\epsilon$, over and above the usual $T$-dependent quantum/thermal
spreading \cite{giordesc,Fulton}. 

This then suggests the enticing possibility that the
 anomalously high value found by Giordano et al., for $T_0$ in Ni wires,
 might be explained by the fluctuating $U_{hyp}$, which allows tunneling
 at a larger bias $\epsilon$ than in the absence of nuclear spins.
 However
 at the present time we do not think this is likely, simply because
 $E_0$
 for this case is too small; as noted immediately following eq.
 (\ref{fracU}), $E_0$ is roughly 2 orders of magnitude smaller than the
 typical barrier height.

 (ii) {\it Resonant Absorption}: In the same way the 
 microwave absorption results \cite{giordres} will have a linewidth
 which
 must be increased by $U_{hyp}(Q)$; the fluctuations in $U_{hyp}$ will
 cause a fluctuation in the resonant frequency, again of order $E_0$,
 and
 again $T$- independent. 
It is interesting to note 
that the linewidths in the Giordano-Hong experiments are rather
large 
(much larger than one would expect from purely dissipative
broadening), 
and in fact the linewidth is roughly what one would expect from
the 
fluctuations in $U_{hyp}(Q)$ for these $Ni$ wires. It would be 
useful to test this explanation of the line broadening. One
obvious
way is to redo the experiments on isotopically purified
samples. 
The effect of the fluctuating $U_{hyp}(Q)$ should also be
discernable as
a low-frequency (of order $\sim ^{\bot}T_{Diff}(\epsilon)$; cf
eq.
(\ref{cond0})) noise; ie., the resonant line is wandering at a
frequency
far lower than microwave frequencies (and far lower than the
frequencies
of thermal and quantum fluctuations). Similar problems are
encountered in
strongly non-linear optical systems, as well on experiments in
glasses.
We emphasize here that a full discussion of the effect of
these dynamic
hyperfine expressions on the experiments is premature; it
requires the
solution of the dynamics of the wall in a time-fluctuating,
spatially random potential. As noted above, this problem is
highly
non-trivial, given the complexity of even the static problem
\cite{Bouchaud}. 
We regard its solution as an important problem- the 
fluctuating $U_{hyp}(Q)$ is so strong in rare earth magnets
that it is 
clear that the domain wall motion will be strongly
influenced by it, up 
to temperatures well above the hyperfine energy. At low
$T$, the domain 
wall will be very strongly pinned by the $U_{hyp}(Q)$,
so much so that 
the low frequency magnetic noise (Barkhausen noise) in
the system will
be completely determined by the time fluctuations in
$U_{hyp}(Q)$. Our 
preliminary investigations of this question indicate
that at low $T$, 
the wall motion will be dominated by "Levy flights",
with a
characteristic anomalous diffusion scaling
behaviour predicted for the 
magnetic noise.

\begin{center}
{\bf ACKNOWLEDGMENTS}
\end{center}

We would like to acknowledge extremely useful discussions with N.
Giordano, K. Hong, and 
N. V. Prokof'ev, and support from NSERC, 
the CIAR, and the UGF fund of the
University of British Columbia.

\newpage

\newpage

\newpage

\newpage

\newpage

\end{document}